\documentclass[preprint,10pt]{elsarticle}

\makeatletter
\def\ps@pprintTitle{%
\let\@oddhead\@empty
\let\@evenhead\@empty
\def\@oddfoot{\centerline{\thepage}}%
\let\@evenfoot\@oddfoot}
\makeatother

\journal{Physica A}

\usepackage{graphicx,bbm,setspace,amssymb,amsfonts,amsmath,mathtools,mathrsfs,stmaryrd,amsthm,bm,etoolbox,comment,hyperref,url,caption,subcaption,color,enumitem,algorithm,booktabs,microtype,nicefrac,lingmacros,nccmath,tree-dvips,tikz-cd}
\usepackage[noend]{algpseudocode}
\usepackage{algorithm}

\usepackage[colorinlistoftodos]{todonotes}
\usepackage[utf8]{inputenc} 
\usepackage[T1]{fontenc}    

\usepackage[noend]{algpseudocode}

\newtheorem{thm-defn}[theorem]{Theorem/Definition}

\theoremstyle{definition}

\theoremstyle{remark}

\DeclareMathOperator{\E}{\mathbb{E}}

\newcommand{\ignore}[1]{}{}

\begin{document}

\begin{frontmatter}

\title{An exploration of the mathematical structure and behavioural biases of 21st century financial crises}
   
\author[label1]{Nick James} \ead{nick.james@unimelb.edu.au}
\author[label2]{Max Menzies} \ead{max.menzies@alumni.harvard.edu}
\address[label1]{School of Mathematics and Statistics, University of Melbourne, Victoria, Australia}
\address[label2]{Yanqi Lake Beijing Institute of Mathematical Sciences and Applications, Beijing, China}

\begin{abstract}
In this paper we contrast the dynamics of the 2022 Ukraine invasion financial crisis with notable financial crises of the 21st century - the dot-com bubble, global financial crisis and COVID-19. We study the similarity in market dynamics and associated implications for equity investors between various financial market crises and we introduce new mathematical techniques to do so. First, we study the strength of collective dynamics during different market crises, and compare suitable portfolio diversification strategies with respect to the unique number of sectors and stocks for optimal systematic risk reduction. Next, we  introduce a new linear operator method to quantify distributional distance between equity returns during various crises. Our method allows us to fairly compare underlying stock and sector performance during different time periods, normalising for those collective dynamics driven by the overall market. Finally, we introduce a new combinatorial portfolio optimisation framework driven by random sampling to investigate whether particular equities and equity sectors are more effective in maximising investor risk-adjusted returns during market crises.     

\end{abstract}


\end{frontmatter}

\section{Introduction}
\label{sec:intro}

Perhaps the most challenging part of investing is making portfolio decisions during economic and financial market crises \cite{Himanshu2021_Covidcrisis}. Most investors and fund managers can make money during bull markets \cite{NYT_bull}, it is bear markets where many investment managers are revealed, as the majority fail to beat index funds \cite{AFR_bear}. Even in academia, a disproportionate amount of research focuses on financial market characteristics that ultimately derive from periods of distress. This includes the identification and modelling of heavy-tailed distributions \cite{Danelsson2006_heavytail,long_statistical_2020}, the identification of regime switching behaviours \cite{Gray1996,Cai1994} and structural breaks \cite{Lamoureux1990,Baillie2009}, optimal portfolio construction \cite{Vercher2007}, heteroskedasticity \cite{Nelson1991}, nonstationarity \cite{Hamilton1989,Livan2012} and perhaps most of all, the nature of heightened volatility \cite{Cerqueti2020,Fang2022}. There is an old saying in financial markets that is often quoted during times of crisis, \textit{``history may not repeat itself, but it does rhyme''} \cite{history_rhyme}. In this work, we address this question mathematically and in a data-driven study - where we introduce new mathematical methods and frameworks to explicitly test the persistence of financial market structure and investor behavioural biases during times of crisis. We are also motivated to compare and contrast the most recent financial crisis brought about by the Russian invasion of Ukraine \cite{rand_ukraine} with prior crises such as the global financial crisis (GFC).

Even if the present day is not strictly speaking a financial crisis, this research is still highly relevant today, given the contemporary market period is quite unprecedented. Following a strong bull market (outside the COVID-19 drawdown and subsequent recovery), we have begun to observe patches of an economic slowdown \cite{deloitte_slowdown}. Many market commentators, investment banks, hedge funds and other high profile stakeholders in the investment business are predicting a sharp economic slowdown and recession, often termed a ``hard landing'' \cite{cnbc_hardlanding}. We are currently experiencing high inflation and money supply around the world \cite{pew_inflation}, and this has triggered central banks to commence structured monetary tightening programs, delivered through systematic increase in interest rates \cite{ap_interestrates}. However, inflation is proving to be sticky and still remains too high according to numerous policymakers \cite{FT_inflation}. Furthermore, we have seen a clear disconnect between fixed income and equity markets, unlike prior financial crises and periods of financial distress \cite{blackrock_disconnect}. To further confuse investors and other financial market participants, we have observed a variety of geopolitical and economic events that fail to align with a singular outlook on market conditions. For instance, the ongoing Ukraine war has had massive ramifications on energy prices \cite{rand_ukraine}, while the technology and financial services sectors have instituted multiple rounds of retrenchments \cite{tech_layoffs}. By contrast, the continued development and hype surrounding generative artificial intelligence potentially brings great opportunity and disruption \cite{McKinsey_AI}. Such changes have caused great contradiction within equity markets. In this work, we explore the differences between the recent financial market period and prior crises. In doing so, we also wish to build on numerous academic work that has analysed financial performance and volatility during market crises, including the Dot-com bubble \cite{Lillo2003}, GFC \cite{Petersen2010}, COVID-19 \cite{Priscilla2022,Lashkaripour2023}, smaller regional crashes \cite{Fauzi2016} and a limited amount of work regarding the Ukraine crisis \cite{Fang2022,Kele2023}. In particular, we seek to investigate the consistency of investor behavioural biases during times of crises, that is, do we see repeat behaviours that can be exploited to generate investor returns.

Our work draws on a long history of applying statistical and physically-inspired models to capture the dynamics of real world phenomena. In financial markets, these methodologes have been applied to a wide range of asset classes such as equities \cite{Wilcox2007,james2022_stagflation,james2021_MJW}, foreign exchange \cite{Ausloos2000} and cryptocurrencies \cite{Gbarowski2019,Wtorek2020,james2021_crypto2,DrodKwapie2022_crypto,DrodWtorek2022_crypto,DrodWtorek2023_crypto,Drod2023_crypto2} and debt-related instruments \cite{Driessen2003}. Such methods from applied mathematics have been used in a variety of other disciplines including epidemiology \cite{jamescovideu,Manchein2020,Li2021_Matjaz,Blasius2020,james2021_TVO,Perc2020,Machado2020,james2021_CovidIndia,Sunahara2023_Matjaz}, environmental sciences \cite{james2022_CO2,Khan2020,Derwent1995,james2021_hydrogen,Westmoreland2007,james2020_Lp,Grange2018,james2023_hydrogen2,Libiseller2005}, crime \cite{james2022_guns,Perc2013,james2023_terrorist}, the arts \cite{Sigaki2018_art,Perc2020_art}, and other fields \cite{james2021_olympics,Clauset2015,james2021_spectral}. Readers interested in recent work related to temporal dynamics with various societal impacts on the economy should consult \cite{Sigaki2019,Perc_social_physics,Perc2019}.

In particular, there is a significant body of work in financial markets that focuses on the cross-correlation matrix of asset price returns as the central object of study. This captures the interdependencies of assets in the market or smaller communities of equities such as GICS sectors. To study the evolution of such correlation structures, techniques such as principal components analysis has been employed, with a particular focus on studying the evolution of the leading eigenvectors and their associated eigenvalue. Various authors have shown that just a few components can describe most of the observed variability of the market
\cite{Pan2007,Fenn2011,Mnnix2012,Heckens2020}. Employing techniques such as random matrix theory, further nuances in the market can be uncovered \cite{Laloux1999,Plerou2002,Gopikrishnan2001}. A wealth of research has also covered network analysis, where the stock is viewed as a complex network and the cross-correlation matrix captures the underlying strength between individual assets, or groups of assets
\cite{Onnela2004,Kim2005}. A wide variety of mathematical and statistical methods have been used to study correlation structures in global equity markets over time, and to identify temporal dependence exhibited by various communities of securities \cite{Drod2001,Drod2018,Drod2019,Drod2020}. The econophysics community have applied such techniques to a wide variety of asset classes. Specific attributes of financial time series, such as their nonstationarity and volatility have led to numerous works exploring methods to capture nuanced effects such as their heavy-tailedness and volatility clustering  \cite{ChiaShangJamesChu1996,Chen2018}. These methods often try to overcome limiting assumptions of the correlation metric, such as its assumption of linearity.

Our final motivating topic of financial portfolio optimisation has been addressed from different perspectives across a variety of research communities. Researchers from fields such as operations research, applied mathematics, statistics, econometrics and more, have tackled the problem of forecasting and optimising financial risk and portfolio selection \cite{Sharpe1966,james2021_portfolio,Calvo2014,Vercher2007,james_arjun,Bhansali2007,Moody2001}. In the econophysics community, various approaches have been used to study methods for reducing market risk, ``best-value'' portfolio construction and evolutionary asset allocation. One benefit in tackling such a problem from an econophysics perspective, is the flexibility in candidate problem solving - where econophysicists can draw on techniques developed in a wide variety of mathematical disciplines.

This paper is structured as follows. In Section \ref{sec:overview_crises}, we outline the specific dates of our financial crises, and comment on the change in the distribution of those equity correlations over time. In Section \ref{sec:crisis_dependent_risk_reduction}, we introduce our random sampling model for the study of evolutionary collective dynamics, and focus on contrasting the benefit in diversification across and within equity sectors in various financial crises. In Section \ref{sec:linear_projection_modelling}, we investigate the similarity in the market performance of varius sectors performance after normalising for different periods - this provides a new approach to directly comparing different economic crises appropriately. We conclude with Section \ref{sec:combinatorial_optimization_portfolio_structure}, where we introduce a new combinatorial optimisation framework for identifying the most frequently occurring equities and equity sectors in top-performing portfolios during times of crisis.

\section{Data and overview of financial crisis periods}
\label{sec:overview_crises}

Throughout this paper, we are interested in the study of collective dynamics over time and within distinct market periods. The core object of study is a collection of equity time series sampled from a variety of GICS sectors across two decades of daily price data. 

We define dates of our economic crises as follows:
\begin{itemize}
    \item \textbf{Dot-com bubble:} 01-03-2000 - 01-10-2002;
    \item \textbf{Global financial crisis (GFC):} 03-01-2007 - 03-05-2010;
    \item \textbf{COVID-19:} 11-03-2020 - 31-08-2020;
    \item \textbf{2022 crash associated with Ukraine:} 03-01-2022 - 13-05-2022.
\end{itemize}

\begin{figure*}
    \centering
    \includegraphics[width=1\textwidth]{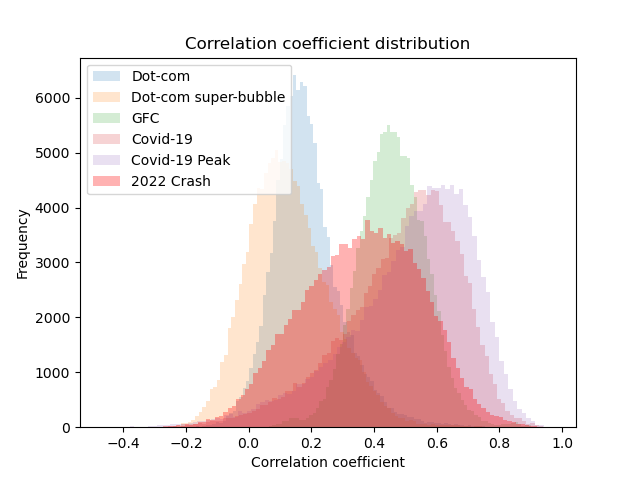}
    \caption{The distribution of equity market correlation coefficients computed in different market crises. Until the Ukraine crisis, there was a clear tendency for equity correlations to become more strongly positive with significantly less variance. The collective severity of correlations decrease among crises in the order COVID-19, the GFC, the Ukraine crash, and the Dot-com.}
    \label{fig:Correlation_coefficient}
\end{figure*}

Figure \ref{fig:Correlation_coefficient} displays the non-trivial correlations between all equities' log returns time series during our periods of investigation. We also include the ``dot-com super-bubble'' \cite{Drod2003_superbubble} (comprising the dot-com bubble up until end of March 2000) and the peak of the COVID-19 financial crisis (11-03-2020 to 29-05-2020). We remark with caution that there may be significant survivorship bias here, particularly with respect to the dot-com bubble (and associated super-bubble) as many stocks that were most significantly affected during this time went bankrupt and do not exist in our analysis. There is a persistent trend from the dot-com bubble until the peak period of the COVID-19 crash, that the average equity correlation during market crises increased quite substantially. This would imply that during crises there is potentially less opportunity for diversification benefit (with less low and negative correlation available), and that economic crises may be shorter and more severe. By contrast, the 2022 market crisis exhibits meaningfully lower average correlation than the GFC, COVID-19 and peak COVID-19 market periods. It is this precise point of difference we use as our springboard for this paper, and seek to investigate the points of difference, and how optimal investor behaviour would deviate during a period such as this, relative to other crises.

\section{Crisis-dependent risk reduction within and across sectors}
\label{sec:crisis_dependent_risk_reduction}

Let $p_i(t)$ be the multivariate time series of equities' daily closing prices where $i=1,...,N$ index the $N=503$ equities under examination. We generate a multivariate time series of log returns as follows:

\begin{equation}
    R_i(t) = \log \left( \frac{p_i(t)}{p_i(t-1)} \right),t=1,...,T.
\end{equation}
We standardise our equity returns via $\tilde{R}_i(t) = [R_i(t) - \langle R_i \rangle ] / \sigma(R_i) $, where $\langle . \rangle $ is an expectation and $\sigma$ represents a standard deviation, both of which are computed over the same time interval. We then choose a a smoothing window $S$ and define a time-varying correlation matrix $\Psi$ as follows:

\begin{equation}
    \Psi(t) = \frac{1}{S} \tilde{R}(t) \tilde{R}^{T}(t),  t= S,...,T.
\end{equation}
There is a wide range of previous literature that focuses on the appropriate value of the parameter $S$ \cite{Fenn2011,james_georg}. when $S$ is too small, changes in the correlation structure will be highly dynamic, however, the results may be excessively noisy. By contrast, when $S$ is large, the evolutionary model may be insensitive to abrupt changes in correlation structure. In this paper, we set $S=60$, shorter than typical values of 90 or 120, as we are investigating shorter and more intense periods of market crisis.

Subsequently, we apply time-varying principal components analysis (PCA) to these time-varying correlation matrices. The key object we examine is the explanatory variance conveyed by the correlation matrix's first eigenvalue. At each instance in time, our evolutionary correlation matrix will generate a sequence of ordered eigenvalues $\lambda_1(t),...,\lambda_N(t)$. We normalise these eigenvalues by  $\tilde{\lambda}_i(t) = \frac{\lambda_i(t)}{\sum^{N}_{k=1} \lambda_k(t)}$.

We now build upon the work first introduced in \cite{james_georg,James2023_cryptoGeorg}, and compare the results from an extensive sampling experiment to highlight differences in across/within equity sector diversification in different market crises. We draw inspiration from the evolutionary collective strength of the market's correlation matrix, measured by $\tilde{\lambda}_1(t)$, to quantify the diversification benefit from various portfolio mixes. For varying values of $(w,a)$ spanning $2 \leq w,a \leq 9$, we draw $D=1000$ random portfolios of $wa$ equities consisting of $a$ sectors and $w$ equities per sector. That is, within each sector (chosen randomly), we sample $w$ equities per sector, and we sample across $a$ sectors. Both the individual securities and equity sectors are drawn randomly and independently with uniform probability.

To quantify diversification benefit for a portfolio that consists of $wa$ equities, we compute the $wa \times wa$ correlation matrix $\bm \Psi$ for each draw and calculate the respective normalised first eigenvalue $\tilde{\lambda}_{w,a}(t)$. We register the 50th percentile (median) of the $D=1000$ values, which we denote $\tilde{\lambda}_{w,a}^{0.50}(t)$. We analyse this quantity in numerous subsequent experiments. We commence by computing the temporal mean of the median of the normalised first eigenvalue, 
\begin{align}
  \mu_{w,a}= \frac{1}{T-\tau+1} \sum_{t=S}^T \tilde{\lambda}_{w,a}^{0.50}(t)
\end{align}
as a measure of the diversification benefit of a portfolio with $w$ equities sampled uniformly (and independently) across $a$ equity GICS sectors, averaged both over time and across our random sampling. We record the values of $\mu_{w,a}$ calculated for each of the four crises under examination in Tables \ref{tab:lambda_1_path_dotcom}, \ref{tab:lambda_1_path_gfc}, \ref{tab:lambda_1_path_covid19} and \ref{tab:lambda_1_path_ukraine}. We term these tables diversification pathways as they offer insight as to how best to increase the collective diversification benefit of a portfolio, by increasing either the number of sectors $a$ or stocks per sectors $w$. Within the tables, we mark in red \emph{greedy paths} that sequentially increase either $w$ or $a$ to maximally decrease the value of $\mu_{w,a}$, that is increase the overall diversification benefit.

We start by examining diversification pathways during the dot-com bubble, shown in Table \ref{tab:lambda_1_path_dotcom}. In exploring the evolution of our greedy strategy, the $\mu_{w,a}$ quantity begins at 0.435 for the (2,2) portfolio and reaches a low-point of 0.221 for the (5,9) portfolio. The greedy path suggests that there is limited further improvement beyond a portfolio consisting of 3 or 4 equities sampled randomly from 5 different equity sectors. The significant reduction in market strength displayed by the average portfolio during the dot-com bubble is representative of decreased correlation among equities compared to either crises (Figure \ref{fig:Correlation_coefficient}), and larger potential to diversify.

We next turn to Table \ref{tab:lambda_1_path_gfc} where we study optimal portfolio construction during the GFC. By contrast, the (2,2) portfolio has a higher level of collective market strength with a $\mu_{w,a} = 0.581$ and only reducing to $\mu_{w,a} = 0.439$ by the time it reaches a (9,6) portfolio. This analysis suggests that during the GFC, there was less opportunity to successfully diversify a portfolio with fewer equities, demonstrated by a significantly larger optimal portfolio.

Optimal portfolio diversification during the COVID-19 market crisis, shown in Table \ref{tab:lambda_1_path_covid19}, commences at the highest point of any crisis we examine. The (2,2) portfolio has a value of $\mu_{w,a}=0.631$, and the optimal portfolio only decreases to a value of $\mu_{w,a}=0.500$, which is achieved by the (9,5) portfolio. There is essentially no  diversification benefit beyond the (6,3) portfolio, which yields a value of $\mu_{w,a} = 0.501$. The limited reduction in average market strength is consistent with the high degree of collective behaviour amongst equities during the COVID-19 market crash and reflects the near impossibility of obtaining a comfortably diversified portfolio during this crisis.

Finally, we turn to the Ukraine market crisis displayed in Table \ref{tab:lambda_1_path_ukraine}. The (2,2) portfolio yields a value of $\mu_{w,a}=0.512$ and reaches a low-point of $\mu_{w,a}=0.351$ with the (5,9) portfolio. This significant reduction is consistent with the clear spread in the distribution of correlation among equities (Figure \ref{fig:Correlation_coefficient}), allowing for substantial reduction in portfolio risk. Furthermore, there appears to be the existence of a reasonable ``best-value'' portfolio, as there is limited reduction in average portfolio risk beyond a portfolio with 3 equities samples from either 5 or 6 equity sectors.

\begin{table*}
\centering
\begin{tabular}{c|rrrrrrrr}
  \hline
 & \multicolumn{8}{c}{Number of equities per sector} \\
  \hline
Number of sectors & 2 & 3 & 4 & 5 & 6 & 7 & 8 & 9 \\ 
  \hline
  2 &\ \textcolor{red}{0.435} & 0.375 & 0.354 & 0.344 & 0.332 & 0.331 & 0.308 & 0.305 \\ 
3 &\ \textcolor{red}{0.341} & 0.320 & 0.294 & 0.290 & 0.284 & 0.280 & 0.268 & 0.274 \\ 
4 &\ \textcolor{red}{0.311} & \textcolor{red}{0.289} & 0.271 & 0.265 & 0.259 & 0.252 & 0.252 & 0.251 \\ 
5 &\ 0.290 & \textcolor{red}{0.268} & \textcolor{red}{0.255} & 0.253 & 0.243 & 0.237 & 0.240 & 0.239 \\ 
6 &\ 0.271 & 0.256 & \textcolor{red}{0.248} & 0.240 & 0.236 & 0.231 & 0.228 & 0.228 \\ 
7 &\ 0.262 & 0.247 & \textcolor{red}{0.237} & 0.234 & 0.231 & 0.230 & 0.224 & 0.226 \\ 
8 &\ 0.248 & 0.240 & \textcolor{red}{0.232} & \textcolor{red}{0.228} & 0.223 & 0.225 & 0.222 & 0.223 \\ 
9 &\ 0.242 & 0.232 & 0.228 & \textcolor{red}{0.221} & 0.221 & 0.220 & 0.218 & 0.216 \\ 
   \hline
\end{tabular}
\caption{Average $\mu_{w,a}$ of the median normalised eigenvalue $\tilde{\lambda}_{w,a}^{0.50}(t)$ for different pairs $(w,a)$ diversifying across $a$ sectors and $w$ equities per sector during the dot-com crisis. In red, we display a greedy path reducing the value of $\mu_{w,a}$ (implying an increase in the overall diversification benefit) by gradually increasing the portfolio size, starting from the smallest portfolio $(2,2)$.
}
\label{tab:lambda_1_path_dotcom}
\end{table*}

\begin{table*}
\centering
\begin{tabular}{c|rrrrrrrr}
  \hline
 & \multicolumn{8}{c}{Number of equities per sector} \\
  \hline
Number of sectors & 2 & 3 & 4 & 5 & 6 & 7 & 8 & 9 \\ 
  \hline
  2 &\ \textcolor{red}{0.581} & 0.550 & 0.530 & 0.526 & 0.510 & 0.499 & 0.505 & 0.497 \\ 
3 &\ \textcolor{red}{0.527} & 0.509 & 0.480 & 0.481 & 0.483 & 0.479 & 0.472 & 0.472 \\ 
4 &\ \textcolor{red}{0.501} & \textcolor{red}{0.479} & 0.475 & 0.470 & 0.464 & 0.460 & 0.456 & 0.456 \\ 
5 &\ 0.482 & \textcolor{red}{0.470} & 0.467 & 0.460 & 0.453 & 0.450 & 0.452 & 0.455 \\ 
6 &\ 0.476 & \textcolor{red}{0.465} & \textcolor{red}{0.451} & 0.453 & 0.448 & 0.445 & 0.448 & 0.444 \\ 
7 &\ 0.461 & 0.454 & \textcolor{red}{0.452} & \textcolor{red}{0.444} & 0.443 & 0.444 & 0.442 & 0.442 \\ 
8 &\ 0.461 & 0.454 & 0.446 & \textcolor{red}{0.442} & \textcolor{red}{0.439} & 0.439 & 0.438 & 0.438 \\ 
9 &\ 0.457 & 0.444 & 0.445 & 0.440 & \textcolor{red}{0.439} & 0.439 & 0.439 & 0.436 \\ 
   \hline
\end{tabular}
\caption{Average $\mu_{w,a}$ of the median normalised eigenvalue $\tilde{\lambda}_{w,a}^{0.50}(t)$ for different pairs $(w,a)$ diversifying across $a$ sectors and $w$ equities per sector during the GFC. In red, we display a greedy path reducing the value of $\mu_{w,a}$ (implying an increase in the overall diversification benefit) by gradually increasing the portfolio size, starting from the smallest portfolio $(2,2)$.}
\label{tab:lambda_1_path_gfc}
\end{table*}

\begin{table*}
\centering
\begin{tabular}{c|rrrrrrrr}
  \hline
 & \multicolumn{8}{c}{Number of equities per sector} \\
  \hline
Number of sectors & 2 & 3 & 4 & 5 & 6 & 7 & 8 & 9 \\ 
  \hline
  2 &\ \textcolor{red}{0.631} & 0.591 & 0.588 & 0.577 & 0.56 & 0.556 & 0.553 & 0.562 \\ 
3 &\ \textcolor{red}{0.578} & 0.566 & 0.543 & 0.543 & 0.543 & 0.553 & 0.538 & 0.531 \\ 
4 &\ \textcolor{red}{0.555} & 0.541 & 0.535 & 0.520 & 0.522 & 0.520 & 0.517 & 0.524 \\ 
5 &\ \textcolor{red}{0.533} & \textcolor{red}{0.516} & 0.520 & 0.501 & 0.517 & 0.515 & 0.504 & 0.504 \\ 
6 &\ 0.522 & \textcolor{red}{0.501} & 0.518 & 0.512 & 0.510 & 0.503 & 0.506 & 0.507 \\ 
7 &\ 0.526 & \textcolor{red}{0.514} & \textcolor{red}{0.507} & \textcolor{red}{0.497} & 0.508 & 0.506 & 0.502 & 0.495 \\ 
8 &\ 0.518 & 0.508 & 0.507 & \textcolor{red}{0.502} & 0.501 & 0.503 & 0.497 & 0.493 \\ 
9 &\ 0.511 & 0.507 & 0.507 & \textcolor{red}{0.500} & 0.502 & 0.502 & 0.501 & 0.495 \\ 
   \hline
\end{tabular}
\caption{Average $\mu_{w,a}$ of the median normalised eigenvalue $\tilde{\lambda}_{w,a}^{0.50}(t)$ for different pairs $(w,a)$ diversifying across $a$ sectors and $w$ equities per sector during the COVID-19 crisis. In red, we display a greedy path reducing the value of $\mu_{w,a}$ (implying an increase in the overall diversification benefit) by gradually increasing the portfolio size, starting from the smallest portfolio $(2,2)$.}
\label{tab:lambda_1_path_covid19}
\end{table*}

\begin{table*}
\centering
\begin{tabular}{c|rrrrrrrr}
  \hline
 & \multicolumn{8}{c}{Number of equities per sector} \\
  \hline
Number of sectors & 2 & 3 & 4 & 5 & 6 & 7 & 8 & 9 \\ 
  \hline
  2 &\ \textcolor{red}{0.512} & 0.489 & 0.467 & 0.473 & 0.469 & 0.466 & 0.447 & 0.444 \\ 
3 &\ \textcolor{red}{0.457} & \textcolor{red}{0.414} & 0.425 & 0.420 & 0.411 & 0.408 & 0.405 & 0.399 \\ 
4 &\ 0.433 & \textcolor{red}{0.402} & 0.393 & 0.395 & 0.393 & 0.381 & 0.377 & 0.379 \\ 
5 &\ 0.408 & \textcolor{red}{0.370} & 0.376 & 0.373 & 0.363 & 0.363 & 0.372 & 0.364 \\ 
6 &\ 0.392 & \textcolor{red}{0.370} & \textcolor{red}{0.365} & 0.368 & 0.368 & 0.364 & 0.353 & 0.357 \\ 
7 &\ 0.391 & 0.369 & \textcolor{red}{0.364} & 0.359 & 0.365 & 0.360 & 0.355 & 0.359 \\ 
8 &\ 0.382 & 0.361 & \textcolor{red}{0.351} & \textcolor{red}{0.353} & 0.353 & 0.350 & 0.353 & 0.348 \\ 
9 &\ 0.378 & 0.361 & 0.358 & \textcolor{red}{0.351} & 0.351 & 0.346 & 0.352 & 0.346 \\ 
   \hline
\end{tabular}
\caption{Average $\mu_{w,a}$ of the median normalised eigenvalue $\tilde{\lambda}_{w,a}^{0.50}(t)$ for different pairs $(w,a)$ diversifying across $a$ sectors and $w$ equities per sector during the 2022 market crisis. In red, we display a greedy path reducing the value of $\mu_{w,a}$ (implying an increase in the overall diversification benefit) by gradually increasing the portfolio size, starting from the smallest portfolio $(2,2)$.}
\label{tab:lambda_1_path_ukraine}
\end{table*}

\begin{figure*}
    \centering
    \includegraphics[width=\textwidth]{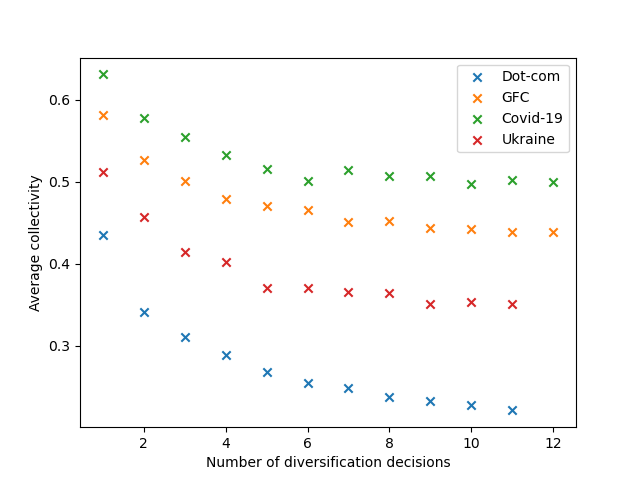}
    \caption{The greedy paths demonstrate the sequential reduction in average collectivity, or increase in collective diversification benefit, as the portfolio increases in size via increases of $w$ or $a$. It is interesting to note the deviation in total risk reduction that is achieved during the dot-com crisis (maximal reduction) when compared to the COVID-19 crisis (minimal reduction). The same order in decreasing collective strength of correlations is observed as in Figure 1: COVID-19, the GFC, the Ukraine crash, and the Dot-com.
    }
    \label{fig:Greedy_paths}
\end{figure*}

To complement the above tables, we directly display the (red) greedy paths in Figure \ref{fig:Greedy_paths}, plotting the value of $\mu_{w,a}$ against the number of steps for each of the four crises. This figure reveals several key findings. First, we see a consistent pattern where there is limited diversification benefit beyond 5 or 6 diversification decisions. Typically this is a portfolio mix of (5,3) or (6,3), where the portfolio is of total size 15 or 18. Together with the tables above, this demonstrates that most diversification steps are made across (rather than within) equity sectors). It is interesting to look at the change in curvature of the this greedy path across various market crises, and how these change after more diversification decisions have been made. Most notably, we see the greatest sequential diversification benefit during the dot-com bubble, and the least during COVID-19 and the GFC. Of particular note is the greedy path that is observed during the COVID-19 crisis, where we observe essentially no increase in diversification potential, just noisy oscillation, beyond step 5. This is consistent with the sharp and abrupt decline in equity markets during this time period, and the indiscriminate selling by various investor types that was observed. By contrast, more protracted crises such as the global financial crisis and the dot-com bubble see continued diversification benefit throughout the course of the respective greedy path. These findings mirror the distributions that are shown in Figure \ref{fig:Correlation_coefficient}, where tighter distributions more strongly translated to the right are consistent with market crises where it is harder to generate diversification benefits.

\begin{figure}
    \centering
\includegraphics[width=\textwidth]{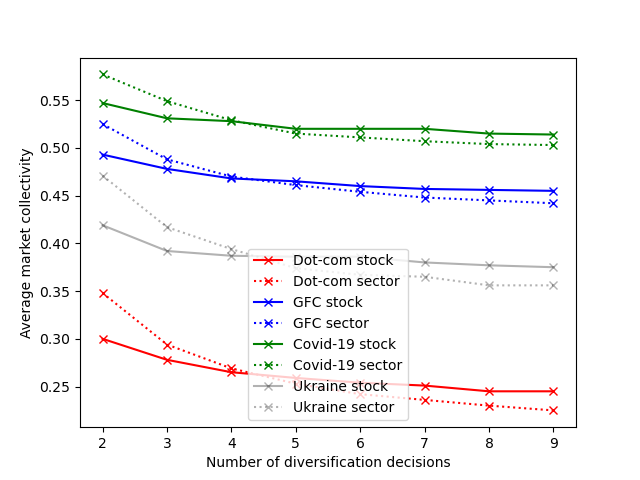}
    \caption{In this figure we contrast the sequential benefit in diversification steps that are taken either across or within equity sectors, having averaged over all possibilities. The paths in the figure depict the change in values of both $\mu_{w,.}=\E_a \mu_{w,a}$ averaged across values of $a$ (a column average of Tables 1-4) and $\mu_{.,a}=\E_w \mu_{w,a}$ averaged across values of $w$ (a row average of Tables 1-4). The figure shows that initially, diversification by stock may provide greater reduction in portfolio risk, however by the time 4 or 5 steps have been taken, there is a clear benefit in subsequent diversification steps taken across sectors, rather than within them. This finding is consistent across all four financial market crises studied. Again, the same order in decreasing collective strength of correlations is observed as in Figure 1: COVID-19, the GFC, the Ukraine crash, and the Dot-com.} 
    \label{fig:Sequential_diversification_lambda_1}
\end{figure}

We conduct one further supporting experiment, where we take an alternative approach to examining the diversification benefit in sequential steps taken across and within equity sectors. In Figure \ref{fig:Sequential_diversification_lambda_1}, we compute and display averaged diversification benefits $\mu_{w,.}=\E_a \mu_{w,a}$ averaged across values of $a$ and  $\mu_{.,a}=\E_w \mu_{w,a}$ averaged across values of $w$. Concretely, these are computed as the column and row averages, respectively, of the above tables of $\mu_{w,a}$, and we display them for each crisis under consideration. In all four market crises, this figure again confirms that beyond 4 or 5 steps, there is greater further reduction in diversifying across sectors, rather than diversifying with different equities sampled from within the same sector. In addition, it again confirms the specific market crises where diversification benefits were most significant. Diversification benefits were most substantial in the dot-com crisis, then Ukraine, the GFC and finally the COVID-19 market crash.

\section{Crisis-to-crisis linear projection modelling}
\label{sec:linear_projection_modelling}

In this section, we start with the following motivating question, \textit{how would equity X have performed in financial period Y?} Here, we introduce a mathematical framework to address this question, and seek to compare the distribution of equity sector returns between financial crises and make inference as to whether sector dynamics or financial crises have greater discriminatory power. For example, do equities from the technology sector behave more similarly in the GFC and dot-com bubble than two different equity sectors (such as consumer discretionary and healthcare) during the same financial crisis. For this purpose, we wish to determine and use an appropriate operator to fairly transform distributions from one period to another. We construct an optimal linear operator as follows.

Given a distribution $f$, there are two natural affine-linear transformations to change it. More simply, there is translation, $f(x) \mapsto h(x)=f(x-b)$ for a constant $b$. Given a distribution, this moves the distribution $f$ $b$ units to the right to yield $h$. In addition, there is scaling, $f(x) \mapsto h(x)=\frac{1}{a} f(\frac{x}{a})$ for $a>0$. The factor of $\frac{1}{a}$ ensures that the transformed $h$ is still a distribution integrating (or summing) to 1. We can combine these two operations to define the following family of linear operators on the space of distributions:
\begin{align}
    T_{a,b} f(x) = \frac{1}{a}f\left(\frac{x-b}{a}\right).
\end{align}
Next, we discuss how to choose the most suitable linear operator to compare two periods. To do this, let $f$ be the distribution of market returns for the GFC, for example, and $g$ be the distribution of market returns for the dot-com. There is unlikely to be a pair $a,b$ such that the transformed distribution $T_{a,b} f$ coincides with $g$ exactly. Instead, we select $a,b$ to minimise the \emph{Wasserstein distance} $d_W(g,T_{a,b} f)$. This will yield a choice $T$ such that $T$ approximately maps returns during the GFC period to returns of the dot-com period.

\begin{figure}
    \centering
    \includegraphics[width=\textwidth]{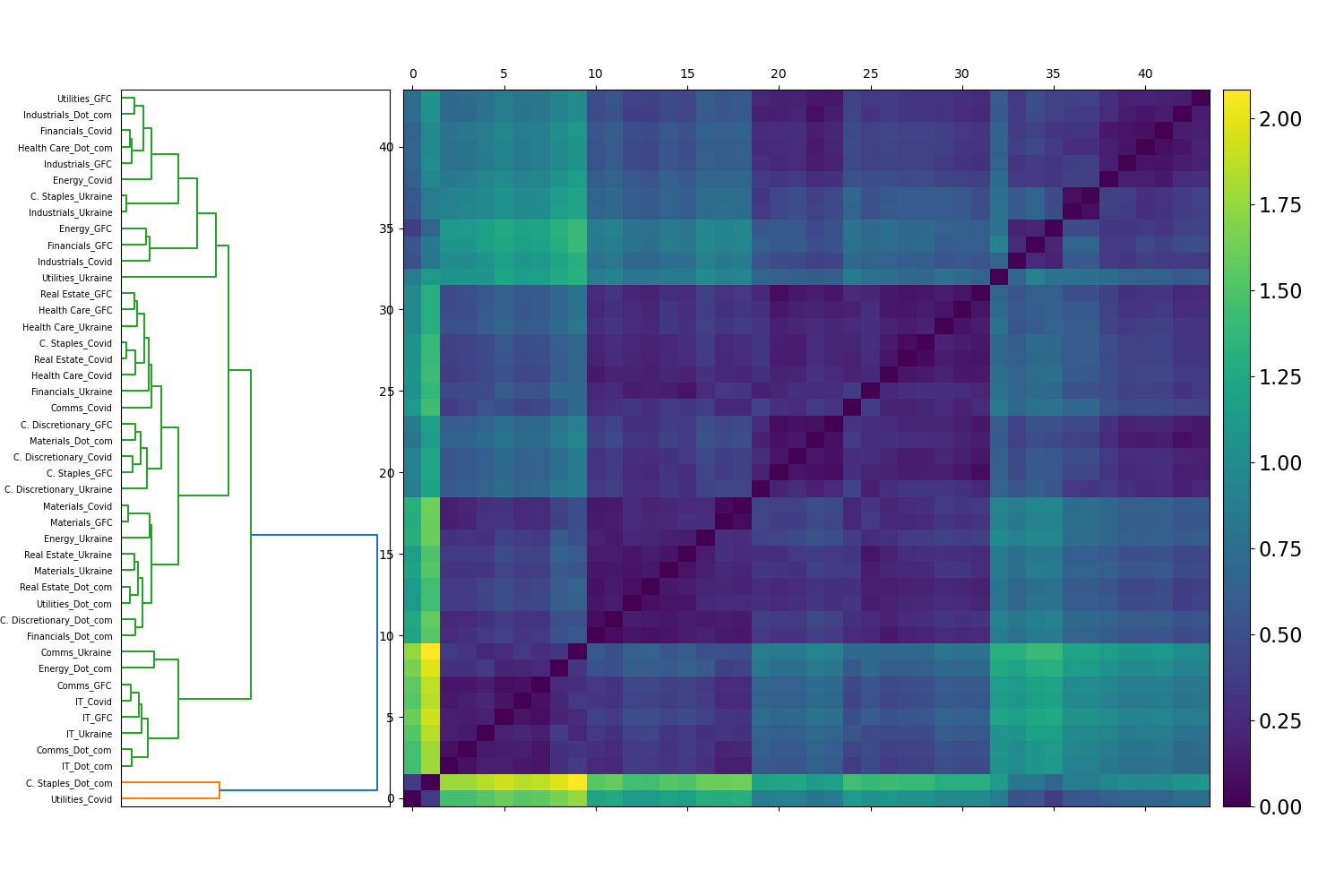}
    \caption{Hierarchical clustering on transformed sector return distributions assesses whether the discriminatory power of individual equity sectors or distinct market crises is stronger. The left y-axis orders and groups different sectors across different time periods according to the similarity of return profiles after normalising for different underlying crisis conditions. The right y-axis gives a colour scale that summarises the size of entries in the distance matrix that forms the bulk of the figure. The cluster structure in the splitting tree of the left y-axis reveals that both individual sectors and market periods exhibit a similar influence in equity market returns. For instance, the information technology sector behaves most similarly with itself across various economic crises while a strong association between the financial, real estate and energy sectors is observed within distinct economic crises.}
    \label{fig:linearoperatordend}
\end{figure}

In Figure \ref{fig:linearoperatordend}, we take every sector $S$ and every crisis $C$ and compare the distributions under the Wasserstein distance. To fairly compare different crises, we use our linear operators determined through our optimisation scheme to map every sector from each crisis to the GFC as a sort of reference period. We explain this in several steps.

First, for notational clarity, we consider each pair of coefficients $a,b$ as already determined and drop them from our notation. Thus, between any two crisis periods $C_1,C_2$ there is an operator $T(C_1,C_2)$. When $C_1=C_2$, this is simply the identity operator. Now, let ${C_0}$ refer to the GFC as our reference crisis period. For each other crisis $C$, we may use our determined linear operator $T({C,C_0})$ to map the returns of $C$ to the GFC $C_0$. Again, for the GFC itself, $T(C,C_0)$ is simply the identity operator, doing nothing. For each sector $S$ of every crisis $C$, we use the comparison operator $T(C,C_0)$ to map $S$ and thus obtain an adjusted distribution of returns $T(C,C_0)S$. Finally, we use the Wasserstein metric to compare all adjusted distributions $T(C,C_0)S$ for each crisis $C$ and sector $S$. Effectively, we have transformed all sectors' return distributions from all different crises into a common metric space where they may be fairly compared. Hierarchical clustering is shown in Figure \ref{fig:linearoperatordend}.

The dendrogram structure in this figure indicates that both individual equity sector idiosyncratic returns and crisis-related impacts exhibit similar discriminatory power. Across the dendrogram one can see multiple instances of association among homogeneous economic crises and equity sectors. Notable self-association can be observed where the same equity sector clusters together when sampled from different economic crises. For example, in a cluster of very high affinity we see strong association between the information technology (IT) and communications sector. This may be due to many equities with a latent IT classification being classified as communications companies under GICS definitions. We see a strong association for energy, real estate and financial sectors to cluster together within any candidate economic crisis. This may be indicative of these sectors' strong relationship with equity beta, and their uniformly strong and weak performance during bull and bear markets, respectively. By contrast, we also see examples of the same equity sector clustering together when sampled from distinct economic crises. For instance, the IT sector returns distribution during the COVID-19 crisis, GFC, Ukraine crisis and dot-com bubble associate together in a small cluster of very high affinity. This indicates that the strength of the IT sector's homogeneity outweighs the collective affinity of distinct economic crises.

\section{Combinatorial simulation for optimal portfolio structure}
\label{sec:combinatorial_optimization_portfolio_structure}

The problem of optimal portfolio weights has been the subject of great debate among financial academics and practitioners for many years. There is a huge range in what various portfolio managers believe is an optimal number of stocks to hold, and the respective weight each of these stocks should comprise within a portfolio. Many quantitative portfolio managers hold hundreds or thousands of equities in a market neutral capacity, seeking to minimise their exposure to the overall market. By contrast, many retail investors or high-conviction fund managers hold as few as 10-15 securities in a long-only capacity - justifying this behaviour as only putting capital behind their very best ideas.

In this section, we begin from the oft-observed assumption that it is difficult to beat equal weightings between assets in portfolio optimisation \cite{DeMiguel2007,Farago2022}. Our philosophy for this section is - when it is so difficult to determine optimal portfolio weights, why even try? Instead, we only vary the selection of assets, using just equal weightings to form our portfolios. We investigate the performance of different stocks (and their composite sectors) in how often they appear in optimised portfolios chosen from random sampling of different assets with equal weightings.

We proceed by performing a portfolio optimisation based on random sampling. We form 100 000 random portfolios, each determined by randomly sampling 40 distinct equities and forming an equally weighted portfolio, with each equity contributing 2.5\% to the overall portfolio. We draw 100 000 random portfolios and extract the top 1\% (or 1000) of these based on ranking their \emph{Sharpe ratios}. Usually the Sharpe ratio optimisation problem takes the following form: one selects weights $w_i$  to optimise the following quantity:

\begin{align}
\label{eq:Sharpeobjectionfn}
 \frac{\sum^{n}_{i=1} w_{i} R_{i} - R_f}{ \sqrt{\boldsymbol{w}^{T} \Sigma \boldsymbol{w}}  }, \\
\text{subject to: } 0 \leq w_{i} \leq 1, i = 1,...,n, \label{constrant1} \\
\sum^{n}_{i=1} w_{i} = 1,\label{constrant2}
\end{align}
where $R_i$ are historical returns, $\Sigma$ the historical covariance matrix, and $R_f$ is the risk-free rate of investment (which we set to zero). This serves as a trade-off between expected portfolio returns and variance.

We make two changes in our setting. First, we reverse the domain of optimisation - $w_i$ are always fixed at $w_i=\frac{1}{40}$ for an equally weighted portfolio, and we optimise only over the choice of assets in the portfolio; second, we sample randomly and deliberately record the best 1\% of portfolios with respect to their Sharpe ratios rather than only the single best. We then study the composition of these top-performing portfolios, and investigate whether this optimal construction varies during different economic crises.

\begin{table}[ht]
\begin{center}
\begin{tabular}{ p{3.6cm}||p{1.65cm}|p{1.35cm}|p{0.9cm}|p{1.1cm}|p{0.8cm}}
  \hline
  Sector & COVID-19 & Dot-com & GFC & Ukraine & Index \\
  \hline
  Energy & 3.8\% & 2.7\% & 2.4\% & 2.6\% & 15.1\% \\ 
  Communications & 3.9\% & 5.1\% & 5.1\% & 6.0\% & 14.1\% \\
  Real Estate & 4.3\% & 6.0\% & 5.4\% & 6.4\% & 13.1\% \\
  Utilities & 5.5\% & 6.9\% & 6.4\% & 8.3\% & 12.7\% \\
  Materials & 6.1\% & 7.5\% & 6.5\% & 8.6\% & 11.5\% \\
  Financials & 9.7\% & 8.2\% & 8.4\% & 9.1\% & 6.6\% \\
  Consumer staples & 10.3\% & 9.1\% & 8.9\% & 10.2\% & 6.2\% \\
  Consumer discretionary & 11.2\% & 12.7\% & 12.7\% & 11.5\% & 5.8\% \\
  Industrials & 14.8\% & 13.5\% & 13.9\% & 11.9\% & 5.6\% \\
  Information technology & 15\% & 13.6\% & 14.5\% & 12.3\% & 5.2\% \\
  Healthcare & 15.4\% & 14.5\% & 16.0\% & 13.1\% & 4.2\% \\ 
  \hline
\end{tabular}
\caption{Sector allocation of top-performing portfolios during crises. For example, among our top 1\% performing portfolios during the COVID-19 market crisis, energy stocks made up 3.8\% across these portfolios, despite comprising 15.1\% of the whole market. We see striking similarity during crises, and significant deviation from the index.}
\label{tab:Simulation_results}
\end{center}
\end{table}

\begin{figure}
    \centering
    \includegraphics[width=\textwidth]{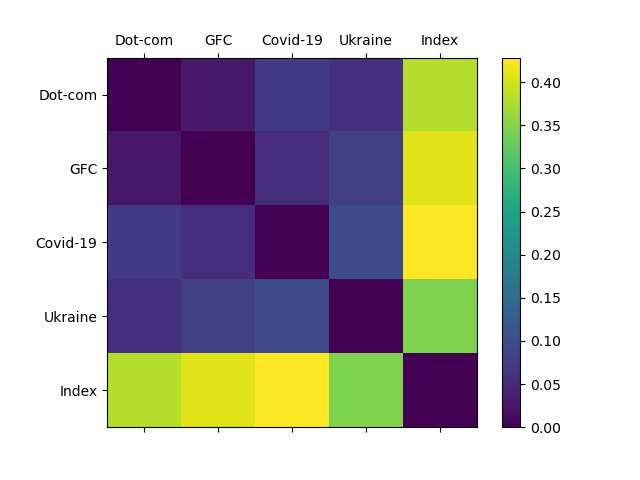}
    \caption{In this figure, we measure the discrepancy between five distributions over equity sectors using the discrete Wasserstein distance (\ref{eq:distributiondist}). The five distributions are the determined distribution of portfolio weights during the four crises, Dot-com, GFC, COVID-19 and Ukraine war and finally the distribution of equity sectors by count. These form the rows and columns of the distance matrix, while the colour scale gives the size of the distance between distributions. The dark square indicates the four distributions of portfolio weights are highly similar (low mutual distance) among all four crises and differ substantially from the distribution by number of sectors. This is seen in Table \ref{tab:Simulation_results} as well, where certain sectors are dramatically over- or under-represented in top-performing portfolios.}
    \label{fig:Discrete_Wasserstein_market_periods}
\end{figure}

For each crisis $C$ and sector $S$, let $p^{(C)}_S$ be the proportion of assets in sector $S$ selected in the optimal random portfolios. These form probability vectors $\mathbf{p}^{(C)} \in \mathbb{R}^{11}$ of length 11,  and constitute the columns of Table \ref{tab:Simulation_results}. Given two crises $C_1,C_2$, we may define a distance between two distributions of chosen sectors as
\begin{align}
\label{eq:distributiondist}
    d(C_1,C_2)= \|\mathbf{p}^{(C_1)} - \mathbf{p}^{(C_2)} \|_1 =\frac12 \sum_{S} |p^{(C_1)}_S - p^{(C_2)}_S|.
\end{align}
where the sum is taken over all 11 sectors $S$. This has the property that $d(C_1,C_2)=0$ if and only if crises $C_1$ and $C_2$ have an identical distribution of stocks chosen across sectors. Furthermore, $d(C_1,C_2) \leq 1$ is the maximal possible distance, with equality if and only if the distributions are disjoint, with no sectors in common at all between the two crises. We also compute the distance between the four crises and the size distribution of sectors according to the count of stocks in each sector (the index column of Table \ref{tab:Simulation_results}.)

While quite simple, this distance may be interpreted as possibly the most suitable distance between two distributions of discrete sets, for it can be shown to be equivalent to the \emph{discrete Wasserstein metric} between distributions on a discrete metric space. Specifically, (\ref{eq:distributiondist}) can be recast in a more general form
\begin{align}
\label{eq:Wassersteinalt}
    W^1 (\mu,\nu) = \sup_{F} \left| \int_{X} F d\mu - \int_{X} F d\nu  \right|.
\end{align}
In (\ref{eq:Wassersteinalt}), $X$ is a discrete set of size 11 equipped with the 0-1 distance, $\mu$ and $\nu$ are the discrete probability distributions associated with $\mathbf{p}^{(C_1)}$ and $\mathbf{p}^{(C_2)}$, respectively, and the supremum is taken over all $1$-Lipschitz functions $F: X \to \mathbb{R}$, meaning \mbox{$|F(x,y)| \leq d(x,y)=1$} for all distinct $x,y$. More details are provided in \cite{James2021_geodesicWasserstein}. We display the matrix of distances between crises $C$ (as well as the index) in Figure \ref{fig:Discrete_Wasserstein_market_periods}.

Table \ref{tab:Simulation_results} shows the frequency of equities sampled from candidate sectors occurring in top-performing portfolios during times of crisis, and reveals numerous surprising findings. First, there is a remarkable consistency of the optimal portfolio membership distribution between market crises, as visible in the four highly similar crisis columns. That is, during the dot-com bubble, GFC, COVID-19 crash and Ukraine war, all periods which exhibited high equity correlations (to varying degrees) and persistent negative returns, the best-performing portfolios displayed incredible similarity in the distribution of sector weightings. Second, and by contrast, the index distribution is meaningfully different to expected optimal portfolio construction during times of crisis. This suggests that during economic crises investors seeking to construct portfolios optimising for risk-adjusted returns will have to deviate materially from index weightings and overweight select equity sectors. The aforementioned two findings are also clearly confirmed by Figure \ref{fig:Discrete_Wasserstein_market_periods}, where under our discrete Wasserstein distance, the index distribution is clearly differentiated from the four crises and their close similarity.

The third finding is the emergence of a clear split in equity sectors that have a tendency to be overweight and underweight. The most predominant top-performing sectors include: financials, consumer staples, consumer discretionary, industrials, information technology and healthcare. Equities from these sectors tend to comprise $\sim 70\% - 75\%$ of total portfolio weight. Equity sectors that occur less frequently include energy, communications, real estate, utilities and materials. These sectors tend to make up  $\sim 20\% - 25\%$ of total portfolio weight. Again, one can see material deviation from optimal weighting during all four crises to the weights one would expect after randomly sampling from the index. It is important to note that the index percentages are based on the number of equity securities existing within each sector and do not reflect each equity's market capitalisation. We are assuming that each security has an equal chance of being held in the portfolio, regardless of other factors such as market capitalisation (which can be highly dependent on recent share price performance).

\section{Conclusion}
\label{sec:Conclusion}

It is worth framing our findings within a broader context of previous econophysics literature, and practical implications for equity investment management. Throughout this paper, we introduce new methodologies and apply them to financial market crises. Although the focal point of this paper is 21st century financial market crises, the methodologies we introduce are flexible and can be generalised well beyond the crises we examine. Concretely, this paper serves two purposes. First, we introduce techniques that could be applied to any time-varying financial system where collective dynamics are of interest (equities, fixed income, currencies, and so on). Second, we apply these techniques on US equity data to learn how equity investors should be managing stock portfolios during times of crisis. We specifically focus on financial market crises as there is greater homogeneity in market dynamics when comparing equity bear markets, than equity bull markets. We have undertaken a detailed and principled study of equity market performance during previous market crises. Our analysis focuses on optimal portfolio construction by way of marginal diversification benefit across and within equity sectors, the discriminatory power and association power of equity sectors and individual market crises, and the most probable portfolio structure to maximise risk-adjusted returns during crises. In each section of our paper, we either extend or develop new mathematical techniques to answer these questions. The methodologies introduced in this paper may be extended quite naturally to any comparative modelling of two candidate financial periods.

The analysis in this paper has direct implications to investors, and their investment decision-making during times of crisis. First, in Section \ref{sec:crisis_dependent_risk_reduction}, where we examine optimal portfolio construction with respect to equities held within and across equity sectors, we demonstrate several interesting findings. Most notably, our analysis identifies a relatively consistent best-value portfolio ranging from 15-18 equities, beyond which the marginal risk reduction from holding an additional equity security is diminished. In the case of the COVID-19 crisis, there is no additional diversification benefit observed at all beyond this point. Our analysis across crises reveals that market crises with higher levels of correlation, unsurprisingly, show materially less diversification benefit as they traverse down the greedy path from (2,2) down to (5,9) or (6,9). In essence, we see surprising consistency in the emergence of a best-value portfolio for risk reduction across all examined market crises. The second key implication to investors comes in Section \ref{sec:combinatorial_optimization_portfolio_structure}, where we introduce a new simulation framework to test for optimal portfolio construction during different market crises. Our framework demonstrates remarkable consistency in the expected optimal portfolio's equity sector weights. Figure \ref{fig:Discrete_Wasserstein_market_periods} crystallises this idea, where the similarity in these weighting schemes is revealed, and perhaps just as importantly, the material deviation from expected index weightings is highlighted. Further, we demonstrate that in the event of a market crisis, implementing successful weighting schemes of prior crises is perhaps the most appropriate strategy. This suggests that equity investors who wish to outperform the index during times of crisis should be guided by successful asset allocation mixes from prior crises. In particular, investors should take note of the fact that select equity sectors such as Healthcare, Information technology, Industrials, Consumer discretionary and Consumer staples are materially overrepresented in optimal portfolios, while sectors such as Energy, Communications, Real estate, Utilities and Materials should be underweighted.

Finally, all the analysis in the paper could be extended beyond US equities. There are many smaller regional financial crises that were not investigated in this paper. Our linear projection analysis (Section \ref{sec:linear_projection_modelling}) demonstrates that both crisis and equity sector homogeneity apply to individual sector performance, to varying degrees. This framework could be extended to the individual security level, and applied more broadly than US equities. This technique is a general approach to answer the question, \textit{``Find a security today that is behaving similarly to how security X behaved in the past.''} This approach could apply to equity and fixed income security selection, and could even be extended to fund manager selection - where we study distributions of hedge fund returns. Future research could investigate equities and other assets more globally to see if there are consistent patterns of sectors or asset classes that are overrepresented in strong-performing portfolios during financial crises or periods of distress more broadly.

\bibliographystyle{_elsarticle-num-names}
\bibliography{__newrefs_fincrises}

\begin{thebibliography}{99}
\expandafter\ifx\csname natexlab\endcsname\relax\def\natexlab#1{#1}\fi
\providecommand{\url}[1]{\texttt{#1}}
\providecommand{\href}[2]{#2}
\providecommand{\path}[1]{#1}
\providecommand{\DOIprefix}{doi:}
\providecommand{\ArXivprefix}{arXiv:}
\providecommand{\URLprefix}{URL: }
\providecommand{\Pubmedprefix}{pmid:}
\providecommand{\doi}[1]{\href{http://dx.doi.org/#1}{\path{#1}}}
\providecommand{\Pubmed}[1]{\href{pmid:#1}{\path{#1}}}
\providecommand{\bibinfo}[2]{#2}
\ifx\xfnm\relax \def\xfnm[#1]{\unskip,\space#1}\fi
\bibitem[{Himanshu et~al.(2021)Himanshu, Ritika, Mushir, and
  Suryavanshi}]{Himanshu2021_Covidcrisis}
\bibinfo{author}{Himanshu}, \bibinfo{author}{Ritika},
  \bibinfo{author}{N.~Mushir}, \bibinfo{author}{R.~Suryavanshi},
\newblock \bibinfo{title}{Impact of {COVID}-19 on portfolio allocation
  decisions of individual investors},
\newblock \bibinfo{journal}{Journal of Public Affairs} \bibinfo{volume}{21}
  (\bibinfo{year}{2021}). \DOIprefix\doi{10.1002/pa.2649}.
\bibitem[{Bernard(2022)}]{NYT_bull}
\bibinfo{author}{T.~S. Bernard}, \bibinfo{title}{Amateur investors rode the
  bull up. now the bear looms},
  \bibinfo{howpublished}{\url{https://www.nytimes.com/2022/05/18/your-money/stock-market-crash-trading-retail.html}},
  \bibinfo{year}{2022}. \bibinfo{note}{{The New York Times}, May 18, 2022}.
\bibitem[{Cain(2020)}]{AFR_bear}
\bibinfo{author}{A.~Cain}, \bibinfo{title}{Index funds outperform active
  managers in a bear market},
  \bibinfo{howpublished}{\url{https://www.afr.com/companies/financial-services/index-funds-outperform-active-managers-in-a-bear-market-20200421-p54lwl}},
  \bibinfo{year}{2020}. \bibinfo{note}{{The Australian Financial Review}, April
  30, 2020}.
\bibitem[{Dan{\'{\i}}elsson et~al.(2006)Dan{\'{\i}}elsson, Jorgensen, Sarma,
  and de~Vries}]{Danelsson2006_heavytail}
\bibinfo{author}{J.~Dan{\'{\i}}elsson}, \bibinfo{author}{B.~N. Jorgensen},
  \bibinfo{author}{M.~Sarma}, \bibinfo{author}{C.~G. de~Vries},
\newblock \bibinfo{title}{Comparing downside risk measures for heavy tailed
  distributions},
\newblock \bibinfo{journal}{Economics Letters} \bibinfo{volume}{92}
  (\bibinfo{year}{2006}) \bibinfo{pages}{202--208}.
  \DOIprefix\doi{10.1016/j.econlet.2006.02.004}.
\bibitem[{Long et~al.(2020)Long, Jebreen, Dassios, and
  Baleanu}]{long_statistical_2020}
\bibinfo{author}{H.~V. Long}, \bibinfo{author}{H.~B. Jebreen},
  \bibinfo{author}{I.~Dassios}, \bibinfo{author}{D.~Baleanu},
\newblock \bibinfo{title}{On the statistical {GARCH} model for managing the
  risk by employing a fat-tailed distribution in finance},
\newblock \bibinfo{journal}{Symmetry} \bibinfo{volume}{12}
  (\bibinfo{year}{2020}) \bibinfo{pages}{1698}.
  \DOIprefix\doi{10.3390/sym12101698}.
\bibitem[{Gray(1996)}]{Gray1996}
\bibinfo{author}{S.~Gray},
\newblock \bibinfo{title}{{Modelling the conditional distribution of interest
  rates as a regime-switching process}},
\newblock \bibinfo{journal}{Journal of Financial Econometrics}
  \bibinfo{volume}{2} (\bibinfo{year}{1996}) \bibinfo{pages}{211--250}.
\bibitem[{Cai(1994)}]{Cai1994}
\bibinfo{author}{J.~Cai},
\newblock \bibinfo{title}{A markov model of switching-regime arch},
\newblock \bibinfo{journal}{Journal of Business \& Economics Statistics}
  \bibinfo{volume}{12} (\bibinfo{year}{1994}) \bibinfo{pages}{309--316}.
\bibitem[{Lamoureux and Lastrapes(1990)}]{Lamoureux1990}
\bibinfo{author}{C.~G. Lamoureux}, \bibinfo{author}{W.~D. Lastrapes},
\newblock \bibinfo{title}{Persistence in variance, structural change, and the
  {GARCH} model},
\newblock \bibinfo{journal}{Journal of Business {\&} Economic Statistics}
  \bibinfo{volume}{8} (\bibinfo{year}{1990}) \bibinfo{pages}{225--234}.
  \DOIprefix\doi{10.2307/1391985}.
\bibitem[{Baillie and Morana(2009)}]{Baillie2009}
\bibinfo{author}{R.~T. Baillie}, \bibinfo{author}{C.~Morana},
\newblock \bibinfo{title}{Modelling long memory and structural breaks in
  conditional variances: An adaptive {FIGARCH} approach},
\newblock \bibinfo{journal}{Journal of Economic Dynamics and Control}
  \bibinfo{volume}{33} (\bibinfo{year}{2009}) \bibinfo{pages}{1577--1592}.
  \DOIprefix\doi{10.1016/j.jedc.2009.02.009}.
\bibitem[{Vercher et~al.(2007)Vercher, Berm{\'{u}}dez, and
  Segura}]{Vercher2007}
\bibinfo{author}{E.~Vercher}, \bibinfo{author}{J.~D. Berm{\'{u}}dez},
  \bibinfo{author}{J.~V. Segura},
\newblock \bibinfo{title}{Fuzzy portfolio optimization under downside risk
  measures},
\newblock \bibinfo{journal}{Fuzzy Sets and Systems} \bibinfo{volume}{158}
  (\bibinfo{year}{2007}) \bibinfo{pages}{769--782}.
  \DOIprefix\doi{10.1016/j.fss.2006.10.026}.
\bibitem[{Nelson(1991)}]{Nelson1991}
\bibinfo{author}{D.~Nelson},
\newblock \bibinfo{title}{Conditional heteroskedasticity in asset returns: A
  new approach},
\newblock \bibinfo{journal}{Econometrica} \bibinfo{volume}{59}
  (\bibinfo{year}{1991}) \bibinfo{pages}{347--370}.
\bibitem[{Hamilton(1989)}]{Hamilton1989}
\bibinfo{author}{J.~D. Hamilton},
\newblock \bibinfo{title}{A new approach to the economic analysis of
  nonstationary time series and the business cycle},
\newblock \bibinfo{journal}{Econometrica} \bibinfo{volume}{57}
  (\bibinfo{year}{1989}) \bibinfo{pages}{357--384}.
  \DOIprefix\doi{10.2307/1912559}.
\bibitem[{Livan et~al.(2012)Livan, ichi Inoue, and Scalas}]{Livan2012}
\bibinfo{author}{G.~Livan}, \bibinfo{author}{J.~ichi Inoue},
  \bibinfo{author}{E.~Scalas},
\newblock \bibinfo{title}{On the non-stationarity of financial time series:
  impact on optimal portfolio selection},
\newblock \bibinfo{journal}{Journal of Statistical Mechanics: Theory and
  Experiment} \bibinfo{volume}{2012} (\bibinfo{year}{2012})
  \bibinfo{pages}{P07025}. \DOIprefix\doi{10.1088/1742-5468/2012/07/p07025}.
\bibitem[{Cerqueti et~al.(2020)Cerqueti, Giacalone, and Mattera}]{Cerqueti2020}
\bibinfo{author}{R.~Cerqueti}, \bibinfo{author}{M.~Giacalone},
  \bibinfo{author}{R.~Mattera},
\newblock \bibinfo{title}{Skewed non-{G}aussian {GARCH} models for
  cryptocurrencies volatility modelling},
\newblock \bibinfo{journal}{Information Sciences} \bibinfo{volume}{527}
  (\bibinfo{year}{2020}) \bibinfo{pages}{1--26}.
  \DOIprefix\doi{10.1016/j.ins.2020.03.075}.
\bibitem[{Fang and Shao(2022)}]{Fang2022}
\bibinfo{author}{Y.~Fang}, \bibinfo{author}{Z.~Shao},
\newblock \bibinfo{title}{The {R}ussia-{U}kraine conflict and volatility risk
  of commodity markets},
\newblock \bibinfo{journal}{Finance Research Letters} \bibinfo{volume}{50}
  (\bibinfo{year}{2022}) \bibinfo{pages}{103264}.
  \DOIprefix\doi{10.1016/j.frl.2022.103264}.
\bibitem[{Salamone(2022)}]{history_rhyme}
\bibinfo{author}{T.~Salamone}, \bibinfo{title}{For stocks, history may not
  repeat itself, but it does rhyme},
  \bibinfo{howpublished}{\url{https://www.nasdaq.com/articles/for-stocks-history-may-not-repeat-itself-but-it-does-rhyme}},
  \bibinfo{year}{2022}. \bibinfo{note}{{Schaeffer's Investment Research}, April
  11, 2022}.
\bibitem[{Jenkins(2023)}]{rand_ukraine}
\bibinfo{author}{B.~M. Jenkins}, \bibinfo{title}{Consequences of the war in
  {U}kraine: The economic fallout},
  \bibinfo{howpublished}{\url{https://www.rand.org/blog/2023/03/consequences-of-the-war-in-ukraine-the-economic-fallout.html}},
  \bibinfo{year}{2023}. \bibinfo{note}{{Rand Corporation}, March 7, 2023}.
\bibitem[{Bachman(2023)}]{deloitte_slowdown}
\bibinfo{author}{D.~Bachman}, \bibinfo{title}{United {S}tates economic
  forecast},
  \bibinfo{howpublished}{\url{https://www2.deloitte.com/us/en/insights/economy/us-economic-forecast/2023-q2.html}},
  \bibinfo{year}{2023}. \bibinfo{note}{{Deloitte}, June 15, 2023}.
\bibitem[{Gilchrist(2023)}]{cnbc_hardlanding}
\bibinfo{author}{K.~Gilchrist}, \bibinfo{title}{{IMF} warns hard landing
  ‘within the realm of possibilities’ for {U.S.} economy},
  \bibinfo{howpublished}{\url{https://www.cnbc.com/2023/04/12/imf-warns-of-hard-landing-risk-for-us-economy.html}},
  \bibinfo{year}{2023}. \bibinfo{note}{{CNBC}, April 12, 2023}.
\bibitem[{DeSilver(2022)}]{pew_inflation}
\bibinfo{author}{D.~DeSilver}, \bibinfo{title}{In the {U.S.} and around the
  world, inflation is high and getting higher},
  \bibinfo{howpublished}{\url{https://www.pewresearch.org/short-reads/2022/06/15/in-the-u-s-and-around-the-world-inflation-is-high-and-getting-higher/}},
  \bibinfo{year}{2022}. \bibinfo{note}{{Pew Research Center}, June 15, 2022}.
\bibitem[{McHUGH(2023)}]{ap_interestrates}
\bibinfo{author}{D.~McHUGH}, \bibinfo{title}{Europe’s central bank hikes key
  interest rate to record high even as recession threat grows},
  \bibinfo{howpublished}{\url{https://apnews.com/article/european-central-bank-interest-rates-inflation-recession-1b7ad9f3f6e9522de05d7b3c1f93e437}},
  \bibinfo{year}{2023}. \bibinfo{note}{{The Associated Press}, September 14,
  2023}.
\bibitem[{Smith and Clarfelt(2023)}]{FT_inflation}
\bibinfo{author}{C.~Smith}, \bibinfo{author}{H.~Clarfelt}, \bibinfo{title}{Jay
  {P}owell warns inflation ‘too high’ in {J}ackson {H}ole speech},
  \bibinfo{howpublished}{\url{https://www.ft.com/content/a1506faf-ced9-4be7-b2f7-2f4c12375786}},
  \bibinfo{year}{2023}. \bibinfo{note}{{The Financial Times}, August 26, 2023}.
\bibitem[{Mathieson(2023)}]{blackrock_disconnect}
\bibinfo{author}{R.~Mathieson}, \bibinfo{title}{Navigating a trilemma},
  \bibinfo{howpublished}{\url{https://www.blackrock.com/us/financial-professionals/insights/systematic-equity-market-outlook}},
  \bibinfo{year}{2023}. \bibinfo{note}{{Blackrock}, April 21, 2023}.
\bibitem[{Stringer(2023)}]{tech_layoffs}
\bibinfo{author}{A.~Stringer}, \bibinfo{title}{A comprehensive list of 2023
  tech layoffs},
  \bibinfo{howpublished}{\url{https://techcrunch.com/2023/09/07/tech-industry-layoffs-2023/}},
  \bibinfo{year}{2023}. \bibinfo{note}{{TechCrunch}, September 7, 2023}.
\bibitem[{McK(2023)}]{McKinsey_AI}
\bibinfo{title}{The economic potential of generative ai: The next productivity
  frontier},
  \bibinfo{howpublished}{\url{https://www.mckinsey.com/capabilities/mckinsey-digital/our-insights/the-economic-potential-of-generative-ai-the-next-productivity-frontier}},
  \bibinfo{year}{2023}. \bibinfo{note}{{McKinsey \& Company}, June 14, 2023}.
\bibitem[{Lillo and Mantegna(2003)}]{Lillo2003}
\bibinfo{author}{F.~Lillo}, \bibinfo{author}{R.~N. Mantegna},
\newblock \bibinfo{title}{Power-law relaxation in a complex system: Omori law
  after a financial market crash},
\newblock \bibinfo{journal}{Physical Review E} \bibinfo{volume}{68}
  (\bibinfo{year}{2003}). \DOIprefix\doi{10.1103/physreve.68.016119}.
\bibitem[{Petersen et~al.(2010)Petersen, Wang, Havlin, and
  Stanley}]{Petersen2010}
\bibinfo{author}{A.~M. Petersen}, \bibinfo{author}{F.~Wang},
  \bibinfo{author}{S.~Havlin}, \bibinfo{author}{H.~E. Stanley},
\newblock \bibinfo{title}{Market dynamics immediately before and after
  financial shocks: Quantifying the omori, productivity, and bath laws},
\newblock \bibinfo{journal}{Physical Review E} \bibinfo{volume}{82}
  (\bibinfo{year}{2010}). \DOIprefix\doi{10.1103/physreve.82.036114}.
\bibitem[{Priscilla et~al.(2022)Priscilla, Hatane, and Tarigan}]{Priscilla2022}
\bibinfo{author}{S.~Priscilla}, \bibinfo{author}{S.~E. Hatane},
  \bibinfo{author}{J.~Tarigan},
\newblock \bibinfo{title}{{COVID}-19 catastrophes and stock market liquidity:
  evidence from technology industry of four biggest {ASEAN} capital market},
\newblock \bibinfo{journal}{Asia-Pacific Journal of Business Administration}
  (\bibinfo{year}{2022}). \DOIprefix\doi{10.1108/apjba-10-2021-0504}.
\bibitem[{Lashkaripour(2023)}]{Lashkaripour2023}
\bibinfo{author}{M.~Lashkaripour},
\newblock \bibinfo{title}{{ESG} tail risk: The {C}ovid-19 market crash
  analysis},
\newblock \bibinfo{journal}{Finance Research Letters} \bibinfo{volume}{53}
  (\bibinfo{year}{2023}) \bibinfo{pages}{103598}.
  \DOIprefix\doi{10.1016/j.frl.2022.103598}.
\bibitem[{Fauzi and Wahyudi(2016)}]{Fauzi2016}
\bibinfo{author}{R.~Fauzi}, \bibinfo{author}{I.~Wahyudi},
\newblock \bibinfo{title}{The effect of firm and stock characteristics on stock
  returns: Stock market crash analysis},
\newblock \bibinfo{journal}{The Journal of Finance and Data Science}
  \bibinfo{volume}{2} (\bibinfo{year}{2016}) \bibinfo{pages}{112--124}.
  \DOIprefix\doi{10.1016/j.jfds.2016.07.001}.
\bibitem[{Kele{\c{s}}(2023)}]{Kele2023}
\bibinfo{author}{E.~Kele{\c{s}}},
\newblock \bibinfo{title}{Stock market response to the {R}ussia-{U}kraine war:
  Evidence from an emerging market},
\newblock \bibinfo{journal}{Journal of East-West Business} \bibinfo{volume}{29}
  (\bibinfo{year}{2023}) \bibinfo{pages}{307--322}.
  \DOIprefix\doi{10.1080/10669868.2023.2210121}.
\bibitem[{Wilcox and Gebbie(2007)}]{Wilcox2007}
\bibinfo{author}{D.~Wilcox}, \bibinfo{author}{T.~Gebbie},
\newblock \bibinfo{title}{An analysis of cross-correlations in an emerging
  market},
\newblock \bibinfo{journal}{Physica A: Statistical Mechanics and its
  Applications} \bibinfo{volume}{375} (\bibinfo{year}{2007})
  \bibinfo{pages}{584--598}. \DOIprefix\doi{10.1016/j.physa.2006.10.030}.
\bibitem[{James et~al.(2022)James, Menzies, and Chin}]{james2022_stagflation}
\bibinfo{author}{N.~James}, \bibinfo{author}{M.~Menzies},
  \bibinfo{author}{K.~Chin},
\newblock \bibinfo{title}{Economic state classification and portfolio
  optimisation with application to stagflationary environments},
\newblock \bibinfo{journal}{Chaos, Solitons \& Fractals} \bibinfo{volume}{164}
  (\bibinfo{year}{2022}) \bibinfo{pages}{112664}.
  \DOIprefix\doi{10.1016/j.chaos.2022.112664}.
\bibitem[{James and Menzies(2021)}]{james2021_MJW}
\bibinfo{author}{N.~James}, \bibinfo{author}{M.~Menzies},
\newblock \bibinfo{title}{A new measure between sets of probability
  distributions with applications to erratic financial behavior},
\newblock \bibinfo{journal}{Journal of Statistical Mechanics: Theory and
  Experiment} \bibinfo{volume}{2021} (\bibinfo{year}{2021})
  \bibinfo{pages}{123404}. \DOIprefix\doi{10.1088/1742-5468/ac3d91}.
\bibitem[{Ausloos(2000)}]{Ausloos2000}
\bibinfo{author}{M.~Ausloos},
\newblock \bibinfo{title}{Statistical physics in foreign exchange currency and
  stock markets},
\newblock \bibinfo{journal}{Physica A: Statistical Mechanics and its
  Applications} \bibinfo{volume}{285} (\bibinfo{year}{2000})
  \bibinfo{pages}{48--65}. \DOIprefix\doi{10.1016/s0378-4371(00)00271-5}.
\bibitem[{G{\k{e}}barowski et~al.(2019)G{\k{e}}barowski,
  O{\'{s}}wi{\k{e}}cimka, W{\k{a}}torek, and
  Dro{\.{z}}d{\.{z}}}]{Gbarowski2019}
\bibinfo{author}{R.~G{\k{e}}barowski},
  \bibinfo{author}{P.~O{\'{s}}wi{\k{e}}cimka},
  \bibinfo{author}{M.~W{\k{a}}torek}, \bibinfo{author}{S.~Dro{\.{z}}d{\.{z}}},
\newblock \bibinfo{title}{Detecting correlations and triangular arbitrage
  opportunities in the forex by means of multifractal detrended
  cross-correlations analysis},
\newblock \bibinfo{journal}{Nonlinear Dynamics} \bibinfo{volume}{98}
  (\bibinfo{year}{2019}) \bibinfo{pages}{2349--2364}.
  \DOIprefix\doi{10.1007/s11071-019-05335-5}.
\bibitem[{W{\k{a}}torek et~al.(2021)W{\k{a}}torek, Dro{\.{z}}d{\.{z}},
  Kwapie{\'{n}}, Minati, O{\'{s}}wi{\k{e}}cimka, and Stanuszek}]{Wtorek2020}
\bibinfo{author}{M.~W{\k{a}}torek}, \bibinfo{author}{S.~Dro{\.{z}}d{\.{z}}},
  \bibinfo{author}{J.~Kwapie{\'{n}}}, \bibinfo{author}{L.~Minati},
  \bibinfo{author}{P.~O{\'{s}}wi{\k{e}}cimka}, \bibinfo{author}{M.~Stanuszek},
\newblock \bibinfo{title}{Multiscale characteristics of the emerging global
  cryptocurrency market},
\newblock \bibinfo{journal}{Physics Reports} \bibinfo{volume}{901}
  (\bibinfo{year}{2021}) \bibinfo{pages}{1--82}.
  \DOIprefix\doi{10.1016/j.physrep.2020.10.005}.
\bibitem[{James and Menzies(2022)}]{james2021_crypto2}
\bibinfo{author}{N.~James}, \bibinfo{author}{M.~Menzies},
\newblock \bibinfo{title}{Collective correlations, dynamics, and behavioural
  inconsistencies of the cryptocurrency market over time},
\newblock \bibinfo{journal}{Nonlinear Dynamics} \bibinfo{volume}{107}
  (\bibinfo{year}{2022}) \bibinfo{pages}{4001--4017}.
  \DOIprefix\doi{10.1007/s11071-021-07166-9}.
\bibitem[{Kwapie{\'{n}} et~al.(2022)Kwapie{\'{n}}, W{\k{a}}torek, Bezbradica,
  Crane, Mai, and Dro{\.{z}}d{\.{z}}}]{DrodKwapie2022_crypto}
\bibinfo{author}{J.~Kwapie{\'{n}}}, \bibinfo{author}{M.~W{\k{a}}torek},
  \bibinfo{author}{M.~Bezbradica}, \bibinfo{author}{M.~Crane},
  \bibinfo{author}{T.~T. Mai}, \bibinfo{author}{S.~Dro{\.{z}}d{\.{z}}},
\newblock \bibinfo{title}{Analysis of inter-transaction time fluctuations in
  the cryptocurrency market},
\newblock \bibinfo{journal}{Chaos: An Interdisciplinary Journal of Nonlinear
  Science} \bibinfo{volume}{32} (\bibinfo{year}{2022}) \bibinfo{pages}{083142}.
  \DOIprefix\doi{10.1063/5.0104707}.
\bibitem[{W{\k{a}}torek et~al.(2022)W{\k{a}}torek, Kwapie{\'{n}}, and
  Dro{\.{z}}d{\.{z}}}]{DrodWtorek2022_crypto}
\bibinfo{author}{M.~W{\k{a}}torek}, \bibinfo{author}{J.~Kwapie{\'{n}}},
  \bibinfo{author}{S.~Dro{\.{z}}d{\.{z}}},
\newblock \bibinfo{title}{Multifractal cross-correlations of bitcoin and ether
  trading characteristics in the post-{COVID}-19 time},
\newblock \bibinfo{journal}{Future Internet} \bibinfo{volume}{14}
  (\bibinfo{year}{2022}) \bibinfo{pages}{215}.
  \DOIprefix\doi{10.3390/fi14070215}.
\bibitem[{W{\k{a}}torek et~al.(2023)W{\k{a}}torek, Kwapie{\'{n}}, and
  Dro{\.{z}}d{\.{z}}}]{DrodWtorek2023_crypto}
\bibinfo{author}{M.~W{\k{a}}torek}, \bibinfo{author}{J.~Kwapie{\'{n}}},
  \bibinfo{author}{S.~Dro{\.{z}}d{\.{z}}},
\newblock \bibinfo{title}{Cryptocurrencies are becoming part of the world
  global financial market},
\newblock \bibinfo{journal}{Entropy} \bibinfo{volume}{25}
  (\bibinfo{year}{2023}) \bibinfo{pages}{377}.
  \DOIprefix\doi{10.3390/e25020377}.
\bibitem[{Dro{\.{z}}d{\.{z}} et~al.(2023)Dro{\.{z}}d{\.{z}}, Kwapie{\'{n}}, and
  W{\k{a}}torek}]{Drod2023_crypto2}
\bibinfo{author}{S.~Dro{\.{z}}d{\.{z}}}, \bibinfo{author}{J.~Kwapie{\'{n}}},
  \bibinfo{author}{M.~W{\k{a}}torek},
\newblock \bibinfo{title}{What is mature and what is still emerging in the
  cryptocurrency market?},
\newblock \bibinfo{journal}{Entropy} \bibinfo{volume}{25}
  (\bibinfo{year}{2023}) \bibinfo{pages}{772}.
  \DOIprefix\doi{10.3390/e25050772}.
\bibitem[{Driessen et~al.(2003)Driessen, Melenberg, and Nijman}]{Driessen2003}
\bibinfo{author}{J.~Driessen}, \bibinfo{author}{B.~Melenberg},
  \bibinfo{author}{T.~Nijman},
\newblock \bibinfo{title}{Common factors in international bond returns},
\newblock \bibinfo{journal}{Journal of International Money and Finance}
  \bibinfo{volume}{22} (\bibinfo{year}{2003}) \bibinfo{pages}{629--656}.
  \DOIprefix\doi{10.1016/s0261-5606(03)00046-9}.
\bibitem[{James et~al.(2021)James, Menzies, and Radchenko}]{jamescovideu}
\bibinfo{author}{N.~James}, \bibinfo{author}{M.~Menzies},
  \bibinfo{author}{P.~Radchenko},
\newblock \bibinfo{title}{{COVID}-19 second wave mortality in {E}urope and the
  {U}nited {S}tates},
\newblock \bibinfo{journal}{Chaos: An Interdisciplinary Journal of Nonlinear
  Science} \bibinfo{volume}{31} (\bibinfo{year}{2021}) \bibinfo{pages}{031105}.
  \DOIprefix\doi{10.1063/5.0041569}.
\bibitem[{Manchein et~al.(2020)Manchein, Brugnago, da~Silva, Mendes, and
  Beims}]{Manchein2020}
\bibinfo{author}{C.~Manchein}, \bibinfo{author}{E.~L. Brugnago},
  \bibinfo{author}{R.~M. da~Silva}, \bibinfo{author}{C.~F.~O. Mendes},
  \bibinfo{author}{M.~W. Beims},
\newblock \bibinfo{title}{Strong correlations between power-law growth of
  {COVID}-19 in four continents and the inefficiency of soft quarantine
  strategies},
\newblock \bibinfo{journal}{Chaos: An Interdisciplinary Journal of Nonlinear
  Science} \bibinfo{volume}{30} (\bibinfo{year}{2020}) \bibinfo{pages}{041102}.
  \DOIprefix\doi{10.1063/5.0009454}.
\bibitem[{Li et~al.(2021)Li, Xu, Song, Wang, and Perc}]{Li2021_Matjaz}
\bibinfo{author}{H.-J. Li}, \bibinfo{author}{W.~Xu}, \bibinfo{author}{S.~Song},
  \bibinfo{author}{W.-X. Wang}, \bibinfo{author}{M.~Perc},
\newblock \bibinfo{title}{The dynamics of epidemic spreading on signed
  networks},
\newblock \bibinfo{journal}{Chaos, Solitons {\&} Fractals}
  \bibinfo{volume}{151} (\bibinfo{year}{2021}) \bibinfo{pages}{111294}.
  \DOIprefix\doi{10.1016/j.chaos.2021.111294}.
\bibitem[{Blasius(2020)}]{Blasius2020}
\bibinfo{author}{B.~Blasius},
\newblock \bibinfo{title}{Power-law distribution in the number of confirmed
  {COVID}-19 cases},
\newblock \bibinfo{journal}{Chaos: An Interdisciplinary Journal of Nonlinear
  Science} \bibinfo{volume}{30} (\bibinfo{year}{2020}) \bibinfo{pages}{093123}.
  \DOIprefix\doi{10.1063/5.0013031}.
\bibitem[{James and Menzies(2022)}]{james2021_TVO}
\bibinfo{author}{N.~James}, \bibinfo{author}{M.~Menzies},
\newblock \bibinfo{title}{Estimating a continuously varying offset between
  multivariate time series with application to {COVID}-19 in the {U}nited
  {S}tates},
\newblock \bibinfo{journal}{The European Physical Journal Special Topics}
  \bibinfo{volume}{231} (\bibinfo{year}{2022}) \bibinfo{pages}{3419--3426}.
  \DOIprefix\doi{10.1140/epjs/s11734-022-00430-y}.
\bibitem[{Perc et~al.(2020)Perc, Miksi{\'{c}}, Slavinec, and
  Sto{\v{z}}er}]{Perc2020}
\bibinfo{author}{M.~Perc}, \bibinfo{author}{N.~G. Miksi{\'{c}}},
  \bibinfo{author}{M.~Slavinec}, \bibinfo{author}{A.~Sto{\v{z}}er},
\newblock \bibinfo{title}{Forecasting {COVID}-19},
\newblock \bibinfo{journal}{Frontiers in Physics} \bibinfo{volume}{8}
  (\bibinfo{year}{2020}) \bibinfo{pages}{127}.
  \DOIprefix\doi{10.3389/fphy.2020.00127}.
\bibitem[{Machado and Lopes(2020)}]{Machado2020}
\bibinfo{author}{J.~A.~T. Machado}, \bibinfo{author}{A.~M. Lopes},
\newblock \bibinfo{title}{Rare and extreme events: the case of {COVID}-19
  pandemic},
\newblock \bibinfo{journal}{Nonlinear Dynamics} \bibinfo{volume}{100}
  (\bibinfo{year}{2020}) \bibinfo{pages}{2953–2972}.
  \DOIprefix\doi{10.1007/s11071-020-05680-w}.
\bibitem[{James et~al.(2022)James, Menzies, and Bondell}]{james2021_CovidIndia}
\bibinfo{author}{N.~James}, \bibinfo{author}{M.~Menzies},
  \bibinfo{author}{H.~Bondell},
\newblock \bibinfo{title}{Comparing the dynamics of {COVID}-19 infection and
  mortality in the {U}nited {S}tates, {I}ndia, and {B}razil},
\newblock \bibinfo{journal}{Physica D: Nonlinear Phenomena}
  \bibinfo{volume}{432} (\bibinfo{year}{2022}) \bibinfo{pages}{133158}.
  \DOIprefix\doi{10.1016/j.physd.2022.133158}.
\bibitem[{Sunahara et~al.(2023)Sunahara, Pessa, Perc, and
  Ribeiro}]{Sunahara2023_Matjaz}
\bibinfo{author}{A.~S. Sunahara}, \bibinfo{author}{A.~A.~B. Pessa},
  \bibinfo{author}{M.~Perc}, \bibinfo{author}{H.~V. Ribeiro},
\newblock \bibinfo{title}{Complexity of the {COVID}-19 pandemic in
  {M}aring{\'{a}}},
\newblock \bibinfo{journal}{Scientific Reports} \bibinfo{volume}{13}
  (\bibinfo{year}{2023}). \DOIprefix\doi{10.1038/s41598-023-39815-x}.
\bibitem[{James and Menzies(2022)}]{james2022_CO2}
\bibinfo{author}{N.~James}, \bibinfo{author}{M.~Menzies},
\newblock \bibinfo{title}{Global and regional changes in carbon dioxide
  emissions: 1970-2019},
\newblock \bibinfo{journal}{Physica A: Statistical Mechanics and its
  Applications} \bibinfo{volume}{608} (\bibinfo{year}{2022})
  \bibinfo{pages}{128302}. \DOIprefix\doi{10.1016/j.physa.2022.128302}.
\bibitem[{Khan et~al.(2020)Khan, Khan, and Rehan}]{Khan2020}
\bibinfo{author}{M.~K. Khan}, \bibinfo{author}{M.~I. Khan},
  \bibinfo{author}{M.~Rehan},
\newblock \bibinfo{title}{The relationship between energy consumption, economic
  growth and carbon dioxide emissions in {P}akistan},
\newblock \bibinfo{journal}{Financial Innovation} \bibinfo{volume}{6}
  (\bibinfo{year}{2020}). \DOIprefix\doi{10.1186/s40854-019-0162-0}.
\bibitem[{Derwent et~al.(1995)Derwent, Middleton, Field, Goldstone, Lester, and
  Perry}]{Derwent1995}
\bibinfo{author}{R.~G. Derwent}, \bibinfo{author}{D.~R. Middleton},
  \bibinfo{author}{R.~A. Field}, \bibinfo{author}{M.~E. Goldstone},
  \bibinfo{author}{J.~N. Lester}, \bibinfo{author}{R.~Perry},
\newblock \bibinfo{title}{Analysis and interpretation of air quality data from
  an urban roadside location in central {L}ondon over the period from {J}uly
  1991 to {J}uly 1992},
\newblock \bibinfo{journal}{Atmospheric Environment} \bibinfo{volume}{29}
  (\bibinfo{year}{1995}) \bibinfo{pages}{923--946}.
  \DOIprefix\doi{10.1016/1352-2310(94)00219-b}.
\bibitem[{James and Menzies(2022)}]{james2021_hydrogen}
\bibinfo{author}{N.~James}, \bibinfo{author}{M.~Menzies},
\newblock \bibinfo{title}{Spatio-temporal trends in the propagation and
  capacity of low-carbon hydrogen projects},
\newblock \bibinfo{journal}{International Journal of Hydrogen Energy}
  \bibinfo{volume}{47} (\bibinfo{year}{2022}) \bibinfo{pages}{16775--16784}.
  \DOIprefix\doi{10.1016/j.ijhydene.2022.03.198}.
\bibitem[{Westmoreland et~al.(2007)Westmoreland, Carslaw, Carslaw, Gillah, and
  Bates}]{Westmoreland2007}
\bibinfo{author}{E.~J. Westmoreland}, \bibinfo{author}{N.~Carslaw},
  \bibinfo{author}{D.~C. Carslaw}, \bibinfo{author}{A.~Gillah},
  \bibinfo{author}{E.~Bates},
\newblock \bibinfo{title}{Analysis of air quality within a street canyon using
  statistical and dispersion modelling techniques},
\newblock \bibinfo{journal}{Atmospheric Environment} \bibinfo{volume}{41}
  (\bibinfo{year}{2007}) \bibinfo{pages}{9195--9205}.
  \DOIprefix\doi{10.1016/j.atmosenv.2007.07.057}.
\bibitem[{James and Menzies(2023)}]{james2020_Lp}
\bibinfo{author}{N.~James}, \bibinfo{author}{M.~Menzies},
\newblock \bibinfo{title}{Equivalence relations and {$L^p$} distances between
  time series with application to the {B}lack {S}ummer {A}ustralian bushfires},
\newblock \bibinfo{journal}{Physica D: Nonlinear Phenomena}
  \bibinfo{volume}{448} (\bibinfo{year}{2023}) \bibinfo{pages}{133693}.
  \DOIprefix\doi{10.1016/j.physd.2023.133693}.
\bibitem[{Grange et~al.(2018)Grange, Carslaw, Lewis, Boleti, and
  Hueglin}]{Grange2018}
\bibinfo{author}{S.~K. Grange}, \bibinfo{author}{D.~C. Carslaw},
  \bibinfo{author}{A.~C. Lewis}, \bibinfo{author}{E.~Boleti},
  \bibinfo{author}{C.~Hueglin},
\newblock \bibinfo{title}{Random forest meteorological normalisation models for
  {S}wiss {PM}$_{10}$ trend analysis},
\newblock \bibinfo{journal}{Atmospheric Chemistry and Physics}
  \bibinfo{volume}{18} (\bibinfo{year}{2018}) \bibinfo{pages}{6223--6239}.
  \DOIprefix\doi{10.5194/acp-18-6223-2018}.
\bibitem[{James and Menzies(2023)}]{james2023_hydrogen2}
\bibinfo{author}{N.~James}, \bibinfo{author}{M.~Menzies},
\newblock \bibinfo{title}{Distributional trends in the generation and end-use
  sector of low-carbon hydrogen plants},
\newblock \bibinfo{journal}{Hydrogen} \bibinfo{volume}{4}
  (\bibinfo{year}{2023}) \bibinfo{pages}{174--189}.
  \DOIprefix\doi{10.3390/hydrogen4010012}.
\bibitem[{Libiseller et~al.(2005)Libiseller, Grimvall, Wald{\'e}n, and
  Saari}]{Libiseller2005}
\bibinfo{author}{C.~Libiseller}, \bibinfo{author}{A.~Grimvall},
  \bibinfo{author}{J.~Wald{\'e}n}, \bibinfo{author}{H.~Saari},
\newblock \bibinfo{title}{Meteorological normalisation and non-parametric
  smoothing for quality assessment and trend analysis of tropospheric ozone
  data},
\newblock \bibinfo{journal}{Environmental Monitoring and Assessment}
  \bibinfo{volume}{100} (\bibinfo{year}{2005}) \bibinfo{pages}{33--52}.
  \DOIprefix\doi{10.1007/s10661-005-7059-2}.
\bibitem[{James and Menzies(2022)}]{james2022_guns}
\bibinfo{author}{N.~James}, \bibinfo{author}{M.~Menzies},
\newblock \bibinfo{title}{Dual-domain analysis of gun violence incidents in the
  {U}nited {S}tates},
\newblock \bibinfo{journal}{Chaos: An Interdisciplinary Journal of Nonlinear
  Science} \bibinfo{volume}{32} (\bibinfo{year}{2022}) \bibinfo{pages}{111101}.
  \DOIprefix\doi{10.1063/5.0120822}.
\bibitem[{Perc et~al.(2013)Perc, Donnay, and Helbing}]{Perc2013}
\bibinfo{author}{M.~Perc}, \bibinfo{author}{K.~Donnay},
  \bibinfo{author}{D.~Helbing},
\newblock \bibinfo{title}{Understanding recurrent crime as system-immanent
  collective behavior},
\newblock \bibinfo{journal}{{PLoS} {ONE}} \bibinfo{volume}{8}
  (\bibinfo{year}{2013}) \bibinfo{pages}{e76063}.
  \DOIprefix\doi{10.1371/journal.pone.0076063}.
\bibitem[{James et~al.(2023)James, Menzies, Chok, Milner, and
  Milner}]{james2023_terrorist}
\bibinfo{author}{N.~James}, \bibinfo{author}{M.~Menzies},
  \bibinfo{author}{J.~Chok}, \bibinfo{author}{A.~Milner},
  \bibinfo{author}{C.~Milner},
\newblock \bibinfo{title}{Geometric persistence and distributional trends in
  worldwide terrorism},
\newblock \bibinfo{journal}{Chaos, Solitons \& Fractals} \bibinfo{volume}{169}
  (\bibinfo{year}{2023}) \bibinfo{pages}{113277}.
  \DOIprefix\doi{10.1016/j.chaos.2023.113277}.
\bibitem[{Sigaki et~al.(2018)Sigaki, Perc, and Ribeiro}]{Sigaki2018_art}
\bibinfo{author}{H.~Y.~D. Sigaki}, \bibinfo{author}{M.~Perc},
  \bibinfo{author}{H.~V. Ribeiro},
\newblock \bibinfo{title}{History of art paintings through the lens of entropy
  and complexity},
\newblock \bibinfo{journal}{Proceedings of the National Academy of Sciences}
  \bibinfo{volume}{115} (\bibinfo{year}{2018}).
  \DOIprefix\doi{10.1073/pnas.1800083115}.
\bibitem[{Perc(2020)}]{Perc2020_art}
\bibinfo{author}{M.~Perc},
\newblock \bibinfo{title}{Beauty in artistic expressions through the eyes of
  networks and physics},
\newblock \bibinfo{journal}{Journal of The Royal Society Interface}
  \bibinfo{volume}{17} (\bibinfo{year}{2020}) \bibinfo{pages}{20190686}.
  \DOIprefix\doi{10.1098/rsif.2019.0686}.
\bibitem[{James et~al.(2022)James, Menzies, and Bondell}]{james2021_olympics}
\bibinfo{author}{N.~James}, \bibinfo{author}{M.~Menzies},
  \bibinfo{author}{H.~Bondell},
\newblock \bibinfo{title}{In search of peak human athletic potential: a
  mathematical investigation},
\newblock \bibinfo{journal}{Chaos: An Interdisciplinary Journal of Nonlinear
  Science} \bibinfo{volume}{32} (\bibinfo{year}{2022}) \bibinfo{pages}{023110}.
  \DOIprefix\doi{10.1063/5.0073141}.
\bibitem[{Clauset et~al.(2015)Clauset, Kogan, and Redner}]{Clauset2015}
\bibinfo{author}{A.~Clauset}, \bibinfo{author}{M.~Kogan},
  \bibinfo{author}{S.~Redner},
\newblock \bibinfo{title}{Safe leads and lead changes in competitive team
  sports},
\newblock \bibinfo{journal}{Physical Review E} \bibinfo{volume}{91}
  (\bibinfo{year}{2015}) \bibinfo{pages}{062815}.
  \DOIprefix\doi{10.1103/physreve.91.062815}.
\bibitem[{James and Menzies(2022)}]{james2021_spectral}
\bibinfo{author}{N.~James}, \bibinfo{author}{M.~Menzies},
\newblock \bibinfo{title}{Optimally adaptive {B}ayesian spectral density
  estimation for stationary and nonstationary processes},
\newblock \bibinfo{journal}{Statistics and Computing} \bibinfo{volume}{32}
  (\bibinfo{year}{2022}) \bibinfo{pages}{45}.
  \DOIprefix\doi{10.1007/s11222-022-10103-4}.
\bibitem[{Sigaki et~al.(2019)Sigaki, Perc, and Ribeiro}]{Sigaki2019}
\bibinfo{author}{H.~Y.~D. Sigaki}, \bibinfo{author}{M.~Perc},
  \bibinfo{author}{H.~V. Ribeiro},
\newblock \bibinfo{title}{Clustering patterns in efficiency and the
  coming-of-age of the cryptocurrency market},
\newblock \bibinfo{journal}{Scientific Reports} \bibinfo{volume}{9}
  (\bibinfo{year}{2019}) \bibinfo{pages}{1440}.
  \DOIprefix\doi{10.1038/s41598-018-37773-3}.
\bibitem[{Jusup et~al.(2022)Jusup, Holme, Kanazawa, Takayasu, Romi{\'{c}},
  Wang, Ge{\v{c}}ek, Lipi{\'{c}}, Podobnik, Wang, Luo, Klanj{\v{s}}{\v{c}}ek,
  Fan, Boccaletti, and Perc}]{Perc_social_physics}
\bibinfo{author}{M.~Jusup}, \bibinfo{author}{P.~Holme},
  \bibinfo{author}{K.~Kanazawa}, \bibinfo{author}{M.~Takayasu},
  \bibinfo{author}{I.~Romi{\'{c}}}, \bibinfo{author}{Z.~Wang},
  \bibinfo{author}{S.~Ge{\v{c}}ek}, \bibinfo{author}{T.~Lipi{\'{c}}},
  \bibinfo{author}{B.~Podobnik}, \bibinfo{author}{L.~Wang},
  \bibinfo{author}{W.~Luo}, \bibinfo{author}{T.~Klanj{\v{s}}{\v{c}}ek},
  \bibinfo{author}{J.~Fan}, \bibinfo{author}{S.~Boccaletti},
  \bibinfo{author}{M.~Perc},
\newblock \bibinfo{title}{Social physics},
\newblock \bibinfo{journal}{Physics Reports} \bibinfo{volume}{948}
  (\bibinfo{year}{2022}) \bibinfo{pages}{1--148}.
  \DOIprefix\doi{10.1016/j.physrep.2021.10.005}.
\bibitem[{Perc(2019)}]{Perc2019}
\bibinfo{author}{M.~Perc},
\newblock \bibinfo{title}{The social physics collective},
\newblock \bibinfo{journal}{Scientific Reports} \bibinfo{volume}{9}
  (\bibinfo{year}{2019}). \DOIprefix\doi{10.1038/s41598-019-53300-4}.
\bibitem[{Pan and Sinha(2007)}]{Pan2007}
\bibinfo{author}{R.~K. Pan}, \bibinfo{author}{S.~Sinha},
\newblock \bibinfo{title}{Collective behavior of stock price movements in an
  emerging market},
\newblock \bibinfo{journal}{Physical Review E} \bibinfo{volume}{76}
  (\bibinfo{year}{2007}). \DOIprefix\doi{10.1103/physreve.76.046116}.
\bibitem[{Fenn et~al.(2011)Fenn, Porter, Williams, McDonald, Johnson, and
  Jones}]{Fenn2011}
\bibinfo{author}{D.~J. Fenn}, \bibinfo{author}{M.~A. Porter},
  \bibinfo{author}{S.~Williams}, \bibinfo{author}{M.~McDonald},
  \bibinfo{author}{N.~F. Johnson}, \bibinfo{author}{N.~S. Jones},
\newblock \bibinfo{title}{Temporal evolution of financial-market correlations},
\newblock \bibinfo{journal}{Physical Review E} \bibinfo{volume}{84}
  (\bibinfo{year}{2011}) \bibinfo{pages}{026109}.
  \DOIprefix\doi{10.1103/physreve.84.026109}.
\bibitem[{M\"{u}nnix et~al.(2012)M\"{u}nnix, Shimada, Sch\"{a}fer, Leyvraz,
  Seligman, Guhr, and Stanley}]{Mnnix2012}
\bibinfo{author}{M.~C. M\"{u}nnix}, \bibinfo{author}{T.~Shimada},
  \bibinfo{author}{R.~Sch\"{a}fer}, \bibinfo{author}{F.~Leyvraz},
  \bibinfo{author}{T.~H. Seligman}, \bibinfo{author}{T.~Guhr},
  \bibinfo{author}{H.~E. Stanley},
\newblock \bibinfo{title}{Identifying states of a financial market},
\newblock \bibinfo{journal}{Scientific Reports} \bibinfo{volume}{2}
  (\bibinfo{year}{2012}). \DOIprefix\doi{10.1038/srep00644}.
\bibitem[{Heckens et~al.(2020)Heckens, Krause, and Guhr}]{Heckens2020}
\bibinfo{author}{A.~J. Heckens}, \bibinfo{author}{S.~M. Krause},
  \bibinfo{author}{T.~Guhr},
\newblock \bibinfo{title}{Uncovering the dynamics of correlation structures
  relative to the collective market motion},
\newblock \bibinfo{journal}{Journal of Statistical Mechanics: Theory and
  Experiment} \bibinfo{volume}{2020} (\bibinfo{year}{2020})
  \bibinfo{pages}{103402}. \DOIprefix\doi{10.1088/1742-5468/abb6e2}.
\bibitem[{Laloux et~al.(1999)Laloux, Cizeau, Bouchaud, and
  Potters}]{Laloux1999}
\bibinfo{author}{L.~Laloux}, \bibinfo{author}{P.~Cizeau},
  \bibinfo{author}{J.-P. Bouchaud}, \bibinfo{author}{M.~Potters},
\newblock \bibinfo{title}{Noise dressing of financial correlation matrices},
\newblock \bibinfo{journal}{Physical Review Letters} \bibinfo{volume}{83}
  (\bibinfo{year}{1999}) \bibinfo{pages}{1467--1470}.
  \DOIprefix\doi{10.1103/physrevlett.83.1467}.
\bibitem[{Plerou et~al.(2002)Plerou, Gopikrishnan, Rosenow, Amaral, Guhr, and
  Stanley}]{Plerou2002}
\bibinfo{author}{V.~Plerou}, \bibinfo{author}{P.~Gopikrishnan},
  \bibinfo{author}{B.~Rosenow}, \bibinfo{author}{L.~A.~N. Amaral},
  \bibinfo{author}{T.~Guhr}, \bibinfo{author}{H.~E. Stanley},
\newblock \bibinfo{title}{Random matrix approach to cross correlations in
  financial data},
\newblock \bibinfo{journal}{Physical Review E} \bibinfo{volume}{65}
  (\bibinfo{year}{2002}). \DOIprefix\doi{10.1103/physreve.65.066126}.
\bibitem[{Gopikrishnan et~al.(2001)Gopikrishnan, Rosenow, Plerou, and
  Stanley}]{Gopikrishnan2001}
\bibinfo{author}{P.~Gopikrishnan}, \bibinfo{author}{B.~Rosenow},
  \bibinfo{author}{V.~Plerou}, \bibinfo{author}{H.~E. Stanley},
\newblock \bibinfo{title}{Quantifying and interpreting collective behavior in
  financial markets},
\newblock \bibinfo{journal}{Physical Review E} \bibinfo{volume}{64}
  (\bibinfo{year}{2001}). \DOIprefix\doi{10.1103/physreve.64.035106}.
\bibitem[{Onnela et~al.(2004)Onnela, Kaski, and Kert{'e}sz}]{Onnela2004}
\bibinfo{author}{J.-P. Onnela}, \bibinfo{author}{K.~Kaski},
  \bibinfo{author}{J.~Kert{'e}sz},
\newblock \bibinfo{title}{Clustering and information in correlation based
  financial networks},
\newblock \bibinfo{journal}{The European Physical Journal B - Condensed Matter}
  \bibinfo{volume}{38} (\bibinfo{year}{2004}) \bibinfo{pages}{353--362}.
  \DOIprefix\doi{10.1140/epjb/e2004-00128-7}.
\bibitem[{Kim and Jeong(2005)}]{Kim2005}
\bibinfo{author}{D.-H. Kim}, \bibinfo{author}{H.~Jeong},
\newblock \bibinfo{title}{Systematic analysis of group identification in stock
  markets},
\newblock \bibinfo{journal}{Physical Review E} \bibinfo{volume}{72}
  (\bibinfo{year}{2005}). \DOIprefix\doi{10.1103/physreve.72.046133}.
\bibitem[{Dro{\.{z}}d{\.{z}} et~al.(2001)Dro{\.{z}}d{\.{z}}, Gr\"{u}mmer, Ruf,
  and Speth}]{Drod2001}
\bibinfo{author}{S.~Dro{\.{z}}d{\.{z}}}, \bibinfo{author}{F.~Gr\"{u}mmer},
  \bibinfo{author}{F.~Ruf}, \bibinfo{author}{J.~Speth},
\newblock \bibinfo{title}{Towards identifying the world stock market
  cross-correlations: {DAX} versus {D}ow {J}ones},
\newblock \bibinfo{journal}{Physica A: Statistical Mechanics and its
  Applications} \bibinfo{volume}{294} (\bibinfo{year}{2001})
  \bibinfo{pages}{226--234}. \DOIprefix\doi{10.1016/s0378-4371(01)00119-4}.
\bibitem[{Dro{\.{z}}d{\.{z}} et~al.(2018)Dro{\.{z}}d{\.{z}}, G{\k{e}}barowski,
  Minati, O{\'{s}}wi{\k{e}}cimka, and W{\k{a}}torek}]{Drod2018}
\bibinfo{author}{S.~Dro{\.{z}}d{\.{z}}}, \bibinfo{author}{R.~G{\k{e}}barowski},
  \bibinfo{author}{L.~Minati}, \bibinfo{author}{P.~O{\'{s}}wi{\k{e}}cimka},
  \bibinfo{author}{M.~W{\k{a}}torek},
\newblock \bibinfo{title}{Bitcoin market route to maturity? {E}vidence from
  return fluctuations, temporal correlations and multiscaling effects},
\newblock \bibinfo{journal}{Chaos: An Interdisciplinary Journal of Nonlinear
  Science} \bibinfo{volume}{28} (\bibinfo{year}{2018}) \bibinfo{pages}{071101}.
  \DOIprefix\doi{10.1063/1.5036517}.
\bibitem[{Dro{\.{z}}d{\.{z}} et~al.(2019)Dro{\.{z}}d{\.{z}}, Minati,
  O{\'{s}}wi{\k{e}}cimka, Stanuszek, and W{\k{a}}torek}]{Drod2019}
\bibinfo{author}{S.~Dro{\.{z}}d{\.{z}}}, \bibinfo{author}{L.~Minati},
  \bibinfo{author}{P.~O{\'{s}}wi{\k{e}}cimka}, \bibinfo{author}{M.~Stanuszek},
  \bibinfo{author}{M.~W{\k{a}}torek},
\newblock \bibinfo{title}{Signatures of the crypto-currency market decoupling
  from the forex},
\newblock \bibinfo{journal}{Future Internet} \bibinfo{volume}{11}
  (\bibinfo{year}{2019}) \bibinfo{pages}{154}.
  \DOIprefix\doi{10.3390/fi11070154}.
\bibitem[{Dro{\.{z}}d{\.{z}} et~al.(2020)Dro{\.{z}}d{\.{z}}, Minati,
  O{\'{s}}wi{\k{e}}cimka, Stanuszek, and W{\k{a}}torek}]{Drod2020}
\bibinfo{author}{S.~Dro{\.{z}}d{\.{z}}}, \bibinfo{author}{L.~Minati},
  \bibinfo{author}{P.~O{\'{s}}wi{\k{e}}cimka}, \bibinfo{author}{M.~Stanuszek},
  \bibinfo{author}{M.~W{\k{a}}torek},
\newblock \bibinfo{title}{Competition of noise and collectivity in global
  cryptocurrency trading: Route to a self-contained market},
\newblock \bibinfo{journal}{Chaos: An Interdisciplinary Journal of Nonlinear
  Science} \bibinfo{volume}{30} (\bibinfo{year}{2020}) \bibinfo{pages}{023122}.
  \DOIprefix\doi{10.1063/1.5139634}.
\bibitem[{Chu et~al.(1996)Chu, Santoni, and Liu}]{ChiaShangJamesChu1996}
\bibinfo{author}{C.-S.~J. Chu}, \bibinfo{author}{G.~J. Santoni},
  \bibinfo{author}{T.~Liu},
\newblock \bibinfo{title}{Stock market volatility and regime shifts in
  returns},
\newblock \bibinfo{journal}{Information Sciences} \bibinfo{volume}{94}
  (\bibinfo{year}{1996}) \bibinfo{pages}{179--190}.
  \DOIprefix\doi{10.1016/0020-0255(96)00117-x}.
\bibitem[{Chen et~al.(2018)Chen, Gan, and Chen}]{Chen2018}
\bibinfo{author}{G.~Chen}, \bibinfo{author}{M.~Gan}, \bibinfo{author}{G.~Chen},
\newblock \bibinfo{title}{Generalized exponential autoregressive models for
  nonlinear time series: Stationarity, estimation and applications},
\newblock \bibinfo{journal}{Information Sciences} \bibinfo{volume}{438}
  (\bibinfo{year}{2018}) \bibinfo{pages}{46--57}.
  \DOIprefix\doi{10.1016/j.ins.2018.01.029}.
\bibitem[{Sharpe(1966)}]{Sharpe1966}
\bibinfo{author}{W.~F. Sharpe},
\newblock \bibinfo{title}{Mutual fund performance},
\newblock \bibinfo{journal}{The Journal of Business} \bibinfo{volume}{39}
  (\bibinfo{year}{1966}) \bibinfo{pages}{119--138}.
  \DOIprefix\doi{10.1086/294846}.
\bibitem[{James et~al.(2023)James, Menzies, and Chan}]{james2021_portfolio}
\bibinfo{author}{N.~James}, \bibinfo{author}{M.~Menzies},
  \bibinfo{author}{J.~Chan},
\newblock \bibinfo{title}{Semi-metric portfolio optimization: a new algorithm
  reducing simultaneous asset shocks},
\newblock \bibinfo{journal}{Econometrics} \bibinfo{volume}{11}
  (\bibinfo{year}{2023}) \bibinfo{pages}{8}.
  \DOIprefix\doi{10.3390/econometrics11010008}.
\bibitem[{Calvo et~al.(2014)Calvo, Ivorra, and Liern}]{Calvo2014}
\bibinfo{author}{C.~Calvo}, \bibinfo{author}{C.~Ivorra},
  \bibinfo{author}{V.~Liern},
\newblock \bibinfo{title}{Fuzzy portfolio selection with non-financial goals:
  exploring the efficient frontier},
\newblock \bibinfo{journal}{Annals of Operations Research}
  \bibinfo{volume}{245} (\bibinfo{year}{2014}) \bibinfo{pages}{31--46}.
  \DOIprefix\doi{10.1007/s10479-014-1561-2}.
\bibitem[{Prakash et~al.(2021)Prakash, James, Menzies, and
  Francis}]{james_arjun}
\bibinfo{author}{A.~Prakash}, \bibinfo{author}{N.~James},
  \bibinfo{author}{M.~Menzies}, \bibinfo{author}{G.~Francis},
\newblock \bibinfo{title}{Structural clustering of volatility regimes for
  dynamic trading strategies},
\newblock \bibinfo{journal}{Applied Mathematical Finance} \bibinfo{volume}{28}
  (\bibinfo{year}{2021}) \bibinfo{pages}{236--274}.
  \DOIprefix\doi{10.1080/1350486x.2021.2007146}.
\bibitem[{Bhansali(2007)}]{Bhansali2007}
\bibinfo{author}{V.~Bhansali},
\newblock \bibinfo{title}{Putting economics (back) into quantitative models},
\newblock \bibinfo{journal}{The Journal of Portfolio Management}
  \bibinfo{volume}{33} (\bibinfo{year}{2007}) \bibinfo{pages}{63--76}.
  \DOIprefix\doi{10.3905/jpm.2007.684755}.
\bibitem[{Moody and Saffell(2001)}]{Moody2001}
\bibinfo{author}{J.~Moody}, \bibinfo{author}{M.~Saffell},
\newblock \bibinfo{title}{Learning to trade via direct reinforcement},
\newblock \bibinfo{journal}{{IEEE} Transactions on Neural Networks}
  \bibinfo{volume}{12} (\bibinfo{year}{2001}) \bibinfo{pages}{875--889}.
  \DOIprefix\doi{10.1109/72.935097}.
\bibitem[{Dro{\.{z}}d{\.{z}} et~al.(2003)Dro{\.{z}}d{\.{z}}, Gr\"{u}mmer, Ruf,
  and Speth}]{Drod2003_superbubble}
\bibinfo{author}{S.~Dro{\.{z}}d{\.{z}}}, \bibinfo{author}{F.~Gr\"{u}mmer},
  \bibinfo{author}{F.~Ruf}, \bibinfo{author}{J.~Speth},
\newblock \bibinfo{title}{Log-periodic self-similarity: an emerging financial
  law?},
\newblock \bibinfo{journal}{Physica A: Statistical Mechanics and its
  Applications} \bibinfo{volume}{324} (\bibinfo{year}{2003})
  \bibinfo{pages}{174--182}. \DOIprefix\doi{10.1016/s0378-4371(02)01848-4}.
\bibitem[{James et~al.(2022)James, Menzies, and Gottwald}]{james_georg}
\bibinfo{author}{N.~James}, \bibinfo{author}{M.~Menzies},
  \bibinfo{author}{G.~A. Gottwald},
\newblock \bibinfo{title}{On financial market correlation structures and
  diversification benefits across and within equity sectors},
\newblock \bibinfo{journal}{Physica A: Statistical Mechanics and its
  Applications} \bibinfo{volume}{604} (\bibinfo{year}{2022})
  \bibinfo{pages}{127682}. \DOIprefix\doi{10.1016/j.physa.2022.127682}.
\bibitem[{James and Menzies(2023)}]{James2023_cryptoGeorg}
\bibinfo{author}{N.~James}, \bibinfo{author}{M.~Menzies},
\newblock \bibinfo{title}{Collective dynamics, diversification and optimal
  portfolio construction for cryptocurrencies},
\newblock \bibinfo{journal}{Entropy} \bibinfo{volume}{25}
  (\bibinfo{year}{2023}) \bibinfo{pages}{931}.
  \DOIprefix\doi{10.3390/e25060931}.
\bibitem[{DeMiguel et~al.(2009)DeMiguel, Garlappi, and Uppal}]{DeMiguel2007}
\bibinfo{author}{V.~DeMiguel}, \bibinfo{author}{L.~Garlappi},
  \bibinfo{author}{R.~Uppal},
\newblock \bibinfo{title}{Optimal versus naive diversification: How inefficient
  is the 1/n portfolio strategy?},
\newblock \bibinfo{journal}{Review of Financial Studies} \bibinfo{volume}{22}
  (\bibinfo{year}{2009}) \bibinfo{pages}{1915--1953}.
  \DOIprefix\doi{10.1093/rfs/hhm075}.
\bibitem[{Farago and Hjalmarsson(2023)}]{Farago2022}
\bibinfo{author}{A.~Farago}, \bibinfo{author}{E.~Hjalmarsson},
\newblock \bibinfo{title}{Small rebalanced portfolios often beat the market
  over long horizons},
\newblock \bibinfo{journal}{The Review of Asset Pricing Studies}
  \bibinfo{volume}{13} (\bibinfo{year}{2023}) \bibinfo{pages}{307--342}.
  \DOIprefix\doi{10.1093/rapstu/raac020}.
\bibitem[{James et~al.(2021)James, Menzies, and
  Bondell}]{James2021_geodesicWasserstein}
\bibinfo{author}{N.~James}, \bibinfo{author}{M.~Menzies},
  \bibinfo{author}{H.~Bondell},
\newblock \bibinfo{title}{Understanding spatial propagation using metric
  geometry with application to the spread of {COVID}-19 in the {U}nited
  {S}tates},
\newblock \bibinfo{journal}{{EPL} (Europhysics Letters)} \bibinfo{volume}{135}
  (\bibinfo{year}{2021}) \bibinfo{pages}{48004}.
  \DOIprefix\doi{10.1209/0295-5075/ac2752}.

\end{thebibliography}
\biboptions{sort&compress}

\end{document}